\documentclass[twocolumn,preprintnumbers,amsmath,amssymb,superscriptaddress,nofootinbib]
{revtex4-2}
\usepackage{graphicx}
\usepackage[normalem]{ulem}
\usepackage[svgnames]{xcolor}
\usepackage{bm}
\usepackage{multirow}
\usepackage{float}
\usepackage{titlesec}
\usepackage[columnwise]{lineno}
\usepackage{float}
\usepackage{titlesec}
\usepackage{gensymb}
\usepackage{physics}

\begin{document}
	
	\title{Magneto-Seebeck effect in bismuth}
	\author{Felix Spathelf}
	\altaffiliation{Present address: Physikalisches Institut, Goethe-Universit\"at Frankfurt, 60438 Frankfurt am Main, Germany}
	\email{spathelf@physik.uni-frankfurt.de}
	\affiliation{LPEM (CNRS-Sorbonne University), ESPCI Paris, PSL University, 75005 Paris, France}
	\affiliation{JEIP, USR 3573 CNRS, Coll\`ege de France, PSL University, 75231 Paris Cedex 05, France}
	
	\author{Beno\^{\i}t Fauqu\'e}
	\affiliation{JEIP, USR 3573 CNRS, Coll\`ege de France, PSL University, 75231 Paris Cedex 05, France}
	\author{Kamran Behnia}
	\affiliation{LPEM (CNRS-Sorbonne University), ESPCI Paris, PSL University, 75005 Paris, France}
	
	\date{\today}
	\begin{abstract} 
		Thermoelectricity was discovered almost two centuries ago in bismuth. The large and negative Seebeck coefficient of this semimetal remains almost flat between 300~K and 100~K. This striking feature can be understood by considering the ratio of electron and hole  mobilities and the evolution of their equal densities with temperature. The large and anisotropic magneto-Seebeck effect in bismuth, on the other hand, has not been understood up to the present day. Here, we report on a systematic study of the thermopower of bismuth from room temperature down to 20~K upon application of a magnetic field of 13.8~T in the binary-bisectrix plane. The amplitude of the Seebeck coefficient depends on the orientation of the magnetic field and the anisotropy changes sign with decreasing temperature. The magneto-Seebeck effect becomes non-monotonic at low temperatures. When the magnetic field is oriented along the binary axis, the Seebeck coefficient is not the same for positive and negative fields. This so-called Umkehr effect arises because the high symmetry axes of the Fermi surface ellipsoids are neither parallel to each other nor to the high symmetry axes of the lattice. The complex evolution of thermopower can be accounted for in a large part of the ($T,B,\Theta$)-space by a model based on semiclassical transport theory and incorporating Landau quantization. The employed energy dependence of the scattering time is compatible with electron-acoustic phonon scattering. We find that the transverse Nernst response plays an important role in setting the amplitude of the longitudinal magneto-Seebeck effect. Furthermore, Landau quantization significantly affects thermoelectricity up to temperatures as high as 120~K.
	\end{abstract}
	\maketitle
	
	\section{Introduction}
	
	Thermoelectricity is both of fundamental interest and technologically promising, because it allows to convert waste heat to useful electric power without moving parts. It was observed for the first time almost two centuries ago by T.~J.~Seebeck in bismuth~\cite{Seebeck1825}. The room temperature Seebeck coefficient is smaller in bismuth than in germanium or silicon. But along the trigonal axis, it is as large as $S \approx 100\ \mu\mathrm{V\,K^{-1}}$, which combined with an electrical resistivity of $\rho\approx 135\ \mu\Omega\,\mathrm{cm}$ and a thermal conductivity of $\kappa \approx 6\ \mathrm{W\,K^{-1}\,m^{-1}}$~\cite{Gallo1963} leads to a thermoelectric figure of merit $ZT=\frac{S^2 T}{\kappa \rho}\approx 0.37$, the largest in the periodic table. Bi-Sb alloys have the largest known thermoelectric figure of merit of any solid at cryogenic temperatures and applying a small magnetic field allows to significantly increase $ZT$ further~\cite{Yim1972}. Despite many investigations over a long period of time~\cite{Seebeck1825,Yim1972,Chandrasekhar1959,Gallo1963,Wolfe1963,Michenaud1970,Uher1974,Lenoir1996,Collaudin2014,Hansen1978,Heremans1979,Mikhail1980,Teramoto2008,Popescu2012,Popescu2012a}, the Seebeck effect of bismuth, especially in presence of a finite magnetic field, is far from being understood up to the present day.
	
	Bismuth has extraordinary electronic properties, which give rise to the large, negative and anisotropic Seebeck coefficient. It is a semimetal, i.e.\ the electron density~$n$ equals the hole density~$p$. At low temperatures, they amount only to $n=p=3.0\cdot10^{-17}\ \mathrm{cm^{-3}}$~\cite{Bhargava1967}, being equivalent to one carrier of each sign per $10^5$ atoms as well as a very small Fermi energy. The very large magnetoresistance reflects the extremely high mobility of the charge carriers, which are ballistic at low temperatures~\cite{Prakash2016,Kang2021}. The Fermi surface consists of one hole pocket with parabolic dispersion and three cigar-shaped electron pockets containing Dirac fermions with an extremely anisotropic band structure, the lowest effective mass being equivalent to approximately $10^{-3}$ bare electron masses~\cite{Zhu2011}. The valley degeneracy of the three electron pockets can be lifted by a magnetic field~\cite{Zhu2011a}. High fields even dry up one or two Fermi seas~\cite{Zhu2017}.
	
	During the last two centuries, the Seebeck effect of bismuth has been studied intensively~\cite{Chandrasekhar1959,Gallo1963,Wolfe1963,Michenaud1970,Yim1972,Uher1974,Collaudin2014}. However, surprisingly, no systematic experimental investigation of the Seebeck effect in magnetic field can be found in literature and the highest magnetic field reported amounts only to 5.5~T~\cite{Uher1974}.
	
	Here, we report on a systematic study of the magneto-Seebeck effect of bismuth from room temperature down to 20~K under a magnetic field of up to 13.8~T in the binary-bisectrix plane. We find that the Seebeck coefficient displays a non-trivial evolution with temperature, magnetic field and the orientation of the magnetic field. To explain the experimental results, we developed a model based on semiclassical transport theory. In doing so, we approximated the well established band structure~\cite{Liu1995} by the Lax model~\cite{Lax1960} to account for the non-parabolicity of the electron bands. The scattering time was treated as in Ref.~\cite{Mikhail1980}, implying an energy dependence compatible with electron-phonon scattering. Phonon drag was not included, since it is relevant only at temperatures below 10~K, which is out of the scope of this work~\cite{Mikhail1980,Uher1978,Issi1979}. Because of the very low Fermi energy, Landau quantization is important in bismuth already at comparably low magnetic fields~\cite{Zhu2011}. Therefore, the semiclassical model was extended as to include the effects of Landau quantization. The goal of our theoretical work was to identify the physical mechanisms playing an important role with regard to the magneto-Seebeck effect of bismuth. This is why we aimed at a model, which is as simple as possible and contains as few unknown parameters as possible, instead of perfect agreement with experimental data. Nevertheless, the model reproduces well the observed behavior in a large part of the ($T,B,\Theta$)-space.
	
	We identify two mechanisms which contribute unexpectedly strongly to the magneto-Seebeck effect of bismuth. Firstly, the transverse Nernst response gives rise to a longitudinal Seebeck voltage via the Hall effect. Secondly, Landau quantization significantly affects the thermopower of bismuth up to temperatures as high as 120~K. These effects could also be relevant to other materials with low carrier concentration.
	
	Our experimental and theoretical results are to be compared with previous theoretical studies of the magneto-Seebeck effect in bismuth~\cite{Hansen1978,Heremans1979,Mikhail1980,Teramoto2008,Popescu2012,Popescu2012a}. Our experimental results disagree with the predictions of Ref.~\cite{Popescu2012}. In contrast, for low magnetic fields, our work confirms the formalism of Mikhail et al.~\cite{Mikhail1980}, which is based on a theoretical framework developed by Heremans and Hansen~\cite{Heremans1979}.
	
	\section{Theory}\label{sec:theory}
	\subsection{General}\label{sec:theory:general}
	
	An electric current $\mathbf{j}$ can be generated not only by an electric field $\mathbf{E}$, but also by a thermal gradient $\boldsymbol\nabla T$. This is expressed by
	\begin{equation}\label{eq:j}
		\mathbf{j} = \hat{\sigma} \mathbf{E} -\hat{\alpha}\boldsymbol\nabla T,
	\end{equation}
	where $\hat{\sigma}$ and $\hat{\alpha}$ are the electrical and the thermoelectric conductivities, respectively~\cite{Behnia2015}. For zero current and diagonal conductivity tensors, equation~\eqref{eq:j} leads to
	\begin{equation}
		S_{zz} = \frac{E_z}{\partial_z T} = \frac{\alpha_{zz}}{\sigma_{zz}},
	\end{equation}
	where the first equation is the definition of the Seebeck coefficient $S_{zz}$. In general, however, the tensorial nature of $\hat{\sigma}$ and $\hat{\alpha}$ has to be taken into account:
	\begin{equation}\label{eq:S_tensorial}
		\hat{S} = \hat{\sigma}^{-1} \hat{\alpha} = \hat{\rho} \hat{\alpha}
	\end{equation}
	$\hat{\rho}$ denotes the electrical resistivity tensor. For a magnetic field $\mathbf{B}$ parallel to the $x$-axis, one obtains
	\begin{equation}\label{eq:S_contributions}
		S_{zz} = \rho_{zz} \alpha_{zz} + \rho_{zy} \alpha_{yz}.
	\end{equation}
	Note that the off-diagonal component of the thermoelectric conductivity $\alpha_{yz}$ is commonly associated with the Nernst effect. In the following, we will refer to the first summand of equation~\eqref{eq:S_contributions} as diagonal or longitudinal, whereas the product of the Hall resistivity $\rho_{zy}$ and $\alpha_{yz}$ will be called off-diagonal or transversal contribution.
	
	Time reversal symmetry implies $\hat{\sigma}(\mathbf{B}) = \hat{\sigma}^\mathsf{T} (-\mathbf{B})$ and $\hat{\alpha}(\mathbf{B}) = \hat{\alpha}^\mathsf{T} (-\mathbf{B})$~\cite{Behnia2015}. Therefore, the diagonal entries of $\hat{\sigma}$, $\hat{\alpha}$ and $\hat{\rho}$ are symmetric functions of $\mathbf{B}$. However, time reversal symmetry is not violated by
	\begin{equation}
		\sigma_{ij} (\mathbf{B}) \neq -\sigma_{ij} (-\mathbf{B})\qquad (i\neq j).
	\end{equation}
	The same is true for $\alpha_{ij}$ and $\rho_{ij}$~\cite{Akgoz1975}. Hence, in equation~\eqref{eq:S_contributions}, the term $\rho_{zz} \alpha_{zz}$ is symmetric in $\mathbf{B}$, whereas there are no restrictions on the symmetry of the term $\rho_{zy} \alpha_{yz}$. Therefore, depending on the crystal symmetry, it can happen that $S_{zz} (\mathbf{B}) \neq S_{zz} (-\mathbf{B})$. This behavior is dubbed Umkehr effect~\cite{Akgoz1975}.

	\subsection{The case of bismuth}
	
	\begin{figure*}
		\centering
		\includegraphics[width=0.76\linewidth]{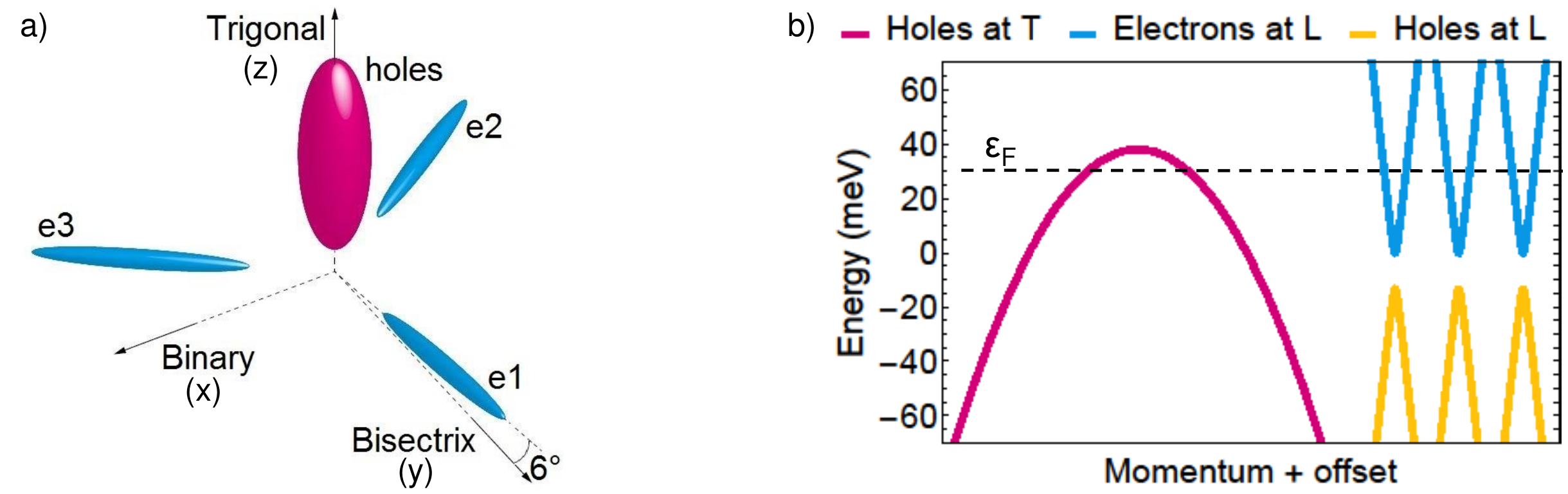}
		\caption{\textbf{Fermi surface and band structure of bismuth:} a) The Fermi surface of bismuth consists of one hole pocket at the $T$-point of the Brillouin zone and three electron pockets at the $L$-points which are tilted by about 6\degree\ out of the binary-bisectrix plane. The Fermi surface is very small as there are only one free electron and one hole per $10^5$ atoms. The Fermi surface is enlarged for better visibility. Adapted from Ref.~\cite{Zhu2018}. b) Dispersion relation of bismuth at 0 K according to the Lax model~\cite{Lax1960}. The valence band at the $T$-point is parabolic, whereas the electrons at the $L$-points have a Dirac-like dispersion due to the energy gap of only 13.6 meV (at zero temperature). Note the very small Fermi energy $\epsilon_F$.}
		\label{fig:Fermi_surface}
	\end{figure*}
	
	In order to calculate the Seebeck coefficient $S_{zz}$, the conductivity tensors $\hat{\sigma}$ and $\hat{\alpha}$ have to be determined. In the case of bismuth, several subtleties of this material have to be considered. In the following, we will focus on the description of these subtleties, whereas a more detailed derivation of the model can be found in the supplement~\cite{Supplement}.
	
	The starting point is the band structure (see Fig.~\ref{fig:Fermi_surface}). The Fermi surface of bismuth consists of one hole pocket at the $T$-point, which is symmetric with respect to the trigonal axis ($z$-axis), and three equivalent electron pockets at the $L$-points of the Brillouin zone. The electron pockets are perpendicular to the binary axis ($x$-axis) and tilted by about 6\degree\ with regard to the bisectrix axis ($y$-axis)~\cite{Zhu2018}. The Fermi surface shows threefold symmetry with the trigonal axis as symmetry axis. This means that there are three binary and three bisectrix axes in the binary-bisectrix (i.e.\ trigonal) plane.
	
	The hole band at the $T$-point has an ordinary parabolic dispersion. In contrast, the dispersion of the electron bands is mostly linear, because at the $L$-points, there is only a very small energy gap $\epsilon_g$ between the conduction and valence bands (see Fig.~\ref{fig:Fermi_surface}b). An appropriate way to describe these Dirac electrons is the Lax model~\cite{Lax1960}:
	\begin{equation}\label{eq:Lax}
		\epsilon (\mathbf{k}) = \pm \frac{1}{2}\left(\epsilon_g^2+2\epsilon_g\hbar^2 \mathbf{k}^\mathsf{T} \hat{m}_{be}^{-1} \mathbf{k}\right)^{1/2}-\frac{1}{2}\epsilon_g
	\end{equation}
	In this context, it is useful to define the quantity $\gamma$ and its derivative with respect to energy~\cite{Heremans1979}:
	\begin{equation}\label{eq:Lax_gamma}
		\gamma (\epsilon) = \epsilon \left(1+\frac{\epsilon}{\epsilon_g}\right) = \frac{\hbar^2}{2} \mathbf{k}^\mathsf{T} \hat{m}_{be}^{-1} \mathbf{k}
	\end{equation}
	\begin{equation}\label{eq:Lax_dgamma}
		\gamma' (\epsilon) = \frac{\partial \gamma}{\partial \epsilon} = 1+2\frac{\epsilon}{\epsilon_g}
	\end{equation}
	In the limit $\epsilon_g \to \infty$, equation~\eqref{eq:Lax} reduces to a quadratic dispersion and can therefore be used to describe the hole band at the $T$-point. In this case, $\gamma$ equals the energy~$\epsilon$ and $\gamma' = 1$.
	
	The inverse mass tensor at the band edge reads
	\begin{equation}\label{eq:inverse_mass_L}
		\hat{m}_{be,L}^{-1}= m_0^{-1}
		\begin{pmatrix}
			806 & 0 & 0\\
			0 & 7.95 & 37.6\\
			0 & 37.6 & 349\\
		\end{pmatrix}
	\end{equation}
	for one of the electron pockets and has to be rotated by 120\degree\ and 240\degree, respectively, around the trigonal axis ($z$-axis) for the other two pockets~\cite{Zhu2011}. $m_0$ denotes the bare electron mass. Note the presence of off-diagonal components in the mass tensor, which is a consequence of the tilt of each electron pocket off the trigonal plane of the crystal~\cite{Brown1968}. This tilt is at the origin of the Umkehr effect in bismuth~\cite{Akgoz1975}.
	
	According to Ref.~\cite{Zhu2011}, the inverse mass of the holes at the $T$-point amounts to
	\begin{equation}\label{eq:inverse_mass_T}
		\hat{m}_{be,T}^{-1}= m_0^{-1}
		\begin{pmatrix}
			14.75 & 0 & 0\\
			0 & 14.75 & 0\\
			0 & 0 & 1.387\\
		\end{pmatrix}.
	\end{equation}
	
	The density of states $D(\epsilon)$ follows from equation~\eqref{eq:Lax}:
	\begin{equation}\label{eq:DOS}
		D(\epsilon) = \sqrt{\frac{2}{\det \hat{m}_{be}^{-1}}} \frac{1}{\pi^2\hbar^3} \sqrt{\gamma (\epsilon)} \gamma'(\epsilon)
	\end{equation}
	
	\begin{figure*}
		\centering
		\includegraphics[width=1\linewidth]{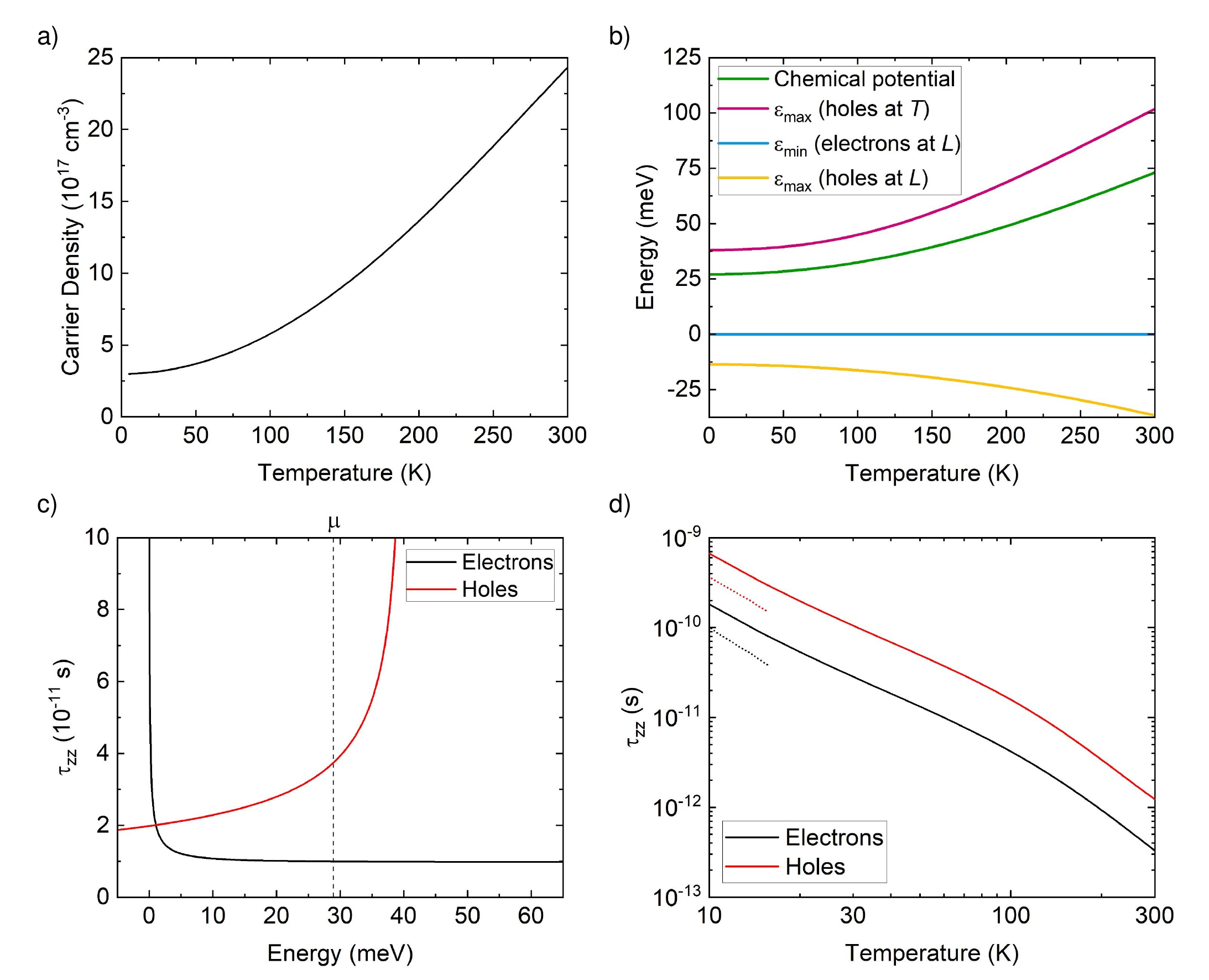}
		\caption{\textbf{Carrier density and band structure vs.\ temperature as well as energy- and temperature-dependence of the scattering time:} a) Carrier density $n=p$ vs.\ temperature as derived from experimental data in Ref.~\cite{Michenaud1972}. b)~Temperature-dependence of the chemical potential $\mu$ and the bottom and top of the electron and hole bands, respectively. c)~Scattering time $\tau_{zz}$ vs.\ energy at $T=60$~K. The divergence at the band edge does not strongly influence the observables, because the density of states is close to zero at these energies. d) Scattering time $\tau_{zz}$ at the chemical potential vs.\ temperature as determined from zero-field resistivity measurements and used for all calculations. For comparison, the dotted lines depict the values found by Hartman~\cite{Hartman1969}.}
		\label{fig:theory}
	\end{figure*}
	
	Unlike other semimetals, the carrier density of bismuth is not constant, but changes by more than a factor of eight between zero and room temperature~(see Fig.~\ref{fig:theory}a)~\cite{Michenaud1972,Issi1979}. This is due to two reasons: Firstly, the thermal broadening of the Fermi-Dirac distribution has a large impact because of the small Fermi energy. Secondly, the band structure is strongly temperature-dependent (see Fig.~\ref{fig:theory}b). For example, the energy gap at the $L$-point almost triples upon heating from 4~K to room temperature~\cite{Vecchi1974,Otake1980}.
	
	In general, the scattering time $\tau$ constitutes the most uncertain ingredient in the description of transport phenomena, because, apart from some proportionalities, it can be directly determined neither theoretically nor experimentally. Here, we assume that it can be described as a product of an energy-dependent scalar $b(\epsilon)$, a temperature-dependent scalar $c(T)$ and a second-order tensor $\hat{a}_p$ capturing the anisotropy, where the index $p$ refers to the $T$- and $L$-points (i.e.\ holes and electrons), respectively:
	\begin{equation}\label{eq:tau_factorization}
		\hat{\tau}_p(\epsilon,T) = \hat{a}_p b(\epsilon) c(T)
	\end{equation}
	Concerning the energy-dependence of the relaxation time, we follow Refs.~\cite{Heremans1979,Mikhail1980,Zawadzki1971,Ravich1971}. Assuming acoustic electron-phonon (and hole-phonon) scattering, one finds Fermi's golden rule
	\begin{equation}\label{eq:scattering_time_general}
		\frac{1}{\tau} \propto D(\epsilon)W^2(\epsilon),
	\end{equation}
	where
	\begin{equation}
		W^2(\epsilon) \propto {\gamma'}^{-2}(\epsilon)
	\end{equation}
	holds for the squared scattering matrix element $W^2$. Inserting these proportionalities into equation~\eqref{eq:tau_factorization} leads to
	\begin{equation}
		\hat{\tau}_p(\epsilon,T) = \hat{a}_p \frac{\gamma'(\epsilon)}{\sqrt{\gamma(\epsilon)}} c(T).
	\end{equation}
	This energy-dependence of $\tau$ is equivalent to an energy-independent mean free path $l$ as predicted for scattering on acoustic phonons\footnote{$l = v\tau = \hbar^{-1} \frac{\partial \epsilon}{\partial k} \tau \propto \frac{\partial \epsilon}{\partial \gamma} \frac{\partial \gamma}{\partial k} \frac{\gamma'(\epsilon)}{\sqrt{\gamma(\epsilon)}} \propto \frac{1}{\gamma'(\epsilon)}\sqrt{\gamma(\epsilon)}\frac{\gamma'(\epsilon)}{\sqrt{\gamma(\epsilon)}} \propto \epsilon^0$}~\cite{Ioffe1960}. For $\epsilon_g \to \infty$, i.e.\ parabolic bands, the energy-dependence reduces as expected to $\tau \propto \epsilon^{-1/2}$~\cite{Heremans2012}. Fig.~\ref{fig:theory}c depicts the scattering time as a function of energy. Once $b(\epsilon)$ is determined, $c(T)$ can be calculated from zero-field resistivity measurements\footnote{Note that $c$ has the dimension M$^{1/2}$L and could be rewritten e.g.\ as $c(T) = \tau_0 (k_B T)^{1/2} \tilde{c}(T)$, where $\tilde{c}(T)$ is a dimensionless function of $T$ and $\tau_0$ a constant with the dimension of time.}. The resulting temperature-dependence of the scattering time is shown in Fig.~\ref{fig:theory}d.
	
	The scattering time $\hat{\tau}_p$ in Eq.~\eqref{eq:tau_factorization} is a tensor. In order to keep time reversal symmetry, it has to be chosen such that $\hat{m}^{-1}_{be,p}\hat{\tau}_p$ is a symmetric tensor~\cite{Mikhail1980}. The five independent variables\footnote{The eight non-zero parameters shown in Eqs.~\eqref{eq:aL} and~\eqref{eq:aT} reduce to five independent parameters, because $a_{T,xx}=a_{T,yy}$ for symmetry reasons, $\hat{m}^{-1}_{be,L}\hat{\tau}_L$ symmetric to keep time reversal symmetry and $a_{L,zz}=1$ as the anisotropy of $\hat{\tau}_p$ is not changed by a factor applied to both $\hat{a}_{T}$ and $\hat{a}_{L}$.} in $\hat{a}_{T}$ and $\hat{a}_{L}$ were the only arbitrary parameters used to adjust the model to all experimental results (Seebeck effect, resistivity and Hall data in the whole accessible $(T,B,\Theta)$-space). The best set of parameters found is
	\begin{equation}\label{eq:aL}
		\hat{a}_{L}=
		\begin{pmatrix}
			0.538 & 0 & 0\\
			0 & 0.610 & -2.64\\
			0 & -0.0180 & 1\\
		\end{pmatrix},
	\end{equation}
	\begin{equation}\label{eq:aT}
		\hat{a}_{T}=
		\begin{pmatrix}
			2.21 & 0 & 0\\
			0 & 2.21 & 0\\
			0 & 0 & 8.04\\
		\end{pmatrix}.
	\end{equation}
	Due to computation time limitations, these values were not determined by an automatic fitting procedure, but by means of a manual heuristic approach with few iterations. Thus, it is very likely that a better agreement between theoretical and experimental curves could be achieved by refining the values of $\hat{a}_{L}$ and $\hat{a}_{T}$. Note that the off-diagonal entries $\hat{a}_{L,23}$ and $\hat{a}_{L,32}$ have to be non-zero in order to fulfill the requirement $\hat{m}^{-1}_{be,L}\hat{\tau}_L = (\hat{m}^{-1}_{be,L}\hat{\tau}_L)^\mathsf{T}$.
	
	In the framework described above, the conductivity tensors of bismuth are given by
	\begin{widetext}
		\begin{equation}\label{eq:sigma}
			\hat{\sigma} = \sum_{\mathrm{pockets}} -\sqrt{\frac{2}{\det \hat{m}_{be}^{-1}}} \frac{2e}{3\pi^2\hbar^3} \int \gamma^{3/2} (\epsilon)
			\left(\left(\frac{e}{\gamma'(\epsilon)}\hat{m}_{be}^{-1}\hat{\tau}(\epsilon,T)\right)^{-1}-\hat{B}\right)^{-1} \frac{\partial f^0}{\partial\epsilon}\ \mathrm{d}\epsilon,
		\end{equation}
		\begin{equation}\label{eq:alpha}
			\hat{\alpha} = \sum_{\mathrm{pockets}} -\sqrt{\frac{2}{\det \hat{m}_{be}^{-1}}} \frac{2}{3\pi^2\hbar^3} \int \frac{\epsilon-\mu}{T} \gamma^{3/2} (\epsilon)
			\left(\left(\frac{e}{\gamma'(\epsilon)}\hat{m}_{be}^{-1}\hat{\tau}(\epsilon,T)\right)^{-1}-\hat{B}\right)^{-1} \frac{\partial f^0}{\partial\epsilon}\ \mathrm{d}\epsilon,
		\end{equation}
	\end{widetext}
	where the sum is taken over the hole pocket at the $T$-point and the three electron pockets at the $L$-points. It was checked that the holes at the $L$-points only contribute negligibly to $\hat{\sigma}$ and $\hat{\alpha}$. $f^0$ and $\mu$ denote the Fermi-Dirac distribution and the chemical potential, respectively. From equations~\eqref{eq:sigma} and~\eqref{eq:alpha}, the zero-field and low-field Seebeck coefficient $S_{zz}$ is determined via equation~\eqref{eq:S_tensorial}. This formalism is equivalent to the one used by Mikhail et al.~\cite{Mikhail1980}. We extended this work by including Landau quantization into the model. In order to do so, the dispersion relation~\eqref{eq:Lax} has to be replaced by
	\begin{widetext}
		\begin{equation}\label{eq:dispersion_Landau}
			\epsilon (j,k_\parallel) = \pm \frac{1}{2}\left(\epsilon_g^2+4\epsilon_g \left(j \hbar \omega_c+\frac{\hbar^2 k_\parallel^2}{2m_{be,\parallel}}\right)\right)^{1/2}-\frac{1}{2}\epsilon_g + g' \mu_B s B.
		\end{equation}
	\end{widetext}
	This equation contains the quantum number $j=n+s+1/2$, where $n \in \mathbb{N}$ and $s=\pm 1/2$ is the spin quantum number~\cite{Wolff1964}, and the cyclotron frequency $\omega_c = eB/m_c$. The term $g' \mu_B s B$ accounts for the effect of the outside bands on spin splitting~\cite{Maltz1970}. The cyclotron mass $m_c$ and the longitudinal effective mass $m_{be,\parallel}$ are calculated from the effective mass tensors according to Ref.~\cite{Zhu2011}, where the values of $g'$ were also taken from. Note that the non-parabolicity of the energy band leads to unequal spacing of the Landau levels.
	
	From equation~\eqref{eq:dispersion_Landau}, the density of states follows as
	\begin{widetext}
		\begin{equation}\label{eq:DOS_Landau}
			D(\epsilon) = \sum_{s=\pm 1/2} \sum_{n=0}^{\infty} \frac{\abs{eB} (2m_{be,\parallel})^{1/2}}{4\pi^2\hbar^2}\frac{\gamma'(\epsilon^\ast)}{(\gamma(\epsilon^\ast) - j\hbar\omega_c)^{1/2}}.
		\end{equation}
	\end{widetext}
	Here, we use $\epsilon^\ast = \epsilon-g' \mu_B s B$ for better readability. As will be shown in section~\ref{sec:results_Landau}, there is an accumulation of electrons in the lowest Landau level leading to a significant change of the carrier density of both electrons and holes in order to keep charge compensation~\cite{Zhu2011,Zhu2017}.
	
	When considering Landau quantization, the conductivity tensors read
	\begin{widetext}
		\begin{equation}\label{eq:sigma_Lax_Landau}
			\hat{\sigma} = \sum_{\mathrm{pockets}} \frac{(2m_{be,\parallel})^{1/2} \abs{eB} e}{2\pi^2\hbar^2} \sum_{s=\pm 1/2} \sum_{n=0}^{\infty} \int (\gamma(\epsilon^\ast)-j\hbar\omega_c)^{1/2} \left(\left(\frac{e}{{\gamma'}(\epsilon^\ast)}\hat{m}_{be}^{-1}\hat{\tau}(\epsilon,T)\right)^{-1}-\hat{B}\right)^{-1} \frac{\partial f^0}{\partial\epsilon}\ \mathrm{d}\epsilon,
		\end{equation}
		\begin{equation}\label{eq:alpha_Lax_Landau}
			\hat{\alpha} = \sum_{\mathrm{pockets}} \frac{(2m_{be,\parallel})^{1/2} \abs{eB}}{2\pi^2\hbar^2} \sum_{s=\pm 1/2} \sum_{n=0}^{\infty} \int \frac{\epsilon-\mu}{T} (\gamma(\epsilon^\ast)-j\hbar\omega_c)^{1/2} \left(\left(\frac{e}{{\gamma'}(\epsilon^\ast)}\hat{m}_{be}^{-1}\hat{\tau}(\epsilon,T)\right)^{-1}-\hat{B}\right)^{-1} \frac{\partial f^0}{\partial\epsilon}\ \mathrm{d}\epsilon.
		\end{equation}
	\end{widetext}
	Refer to the supplement for a more detailed derivation and a discussion of the scattering time in presence of Landau quantization~\cite{Supplement}.

	\section{Experimental details}
	
	\begin{figure}
		\centering
		\includegraphics[width=1\linewidth]{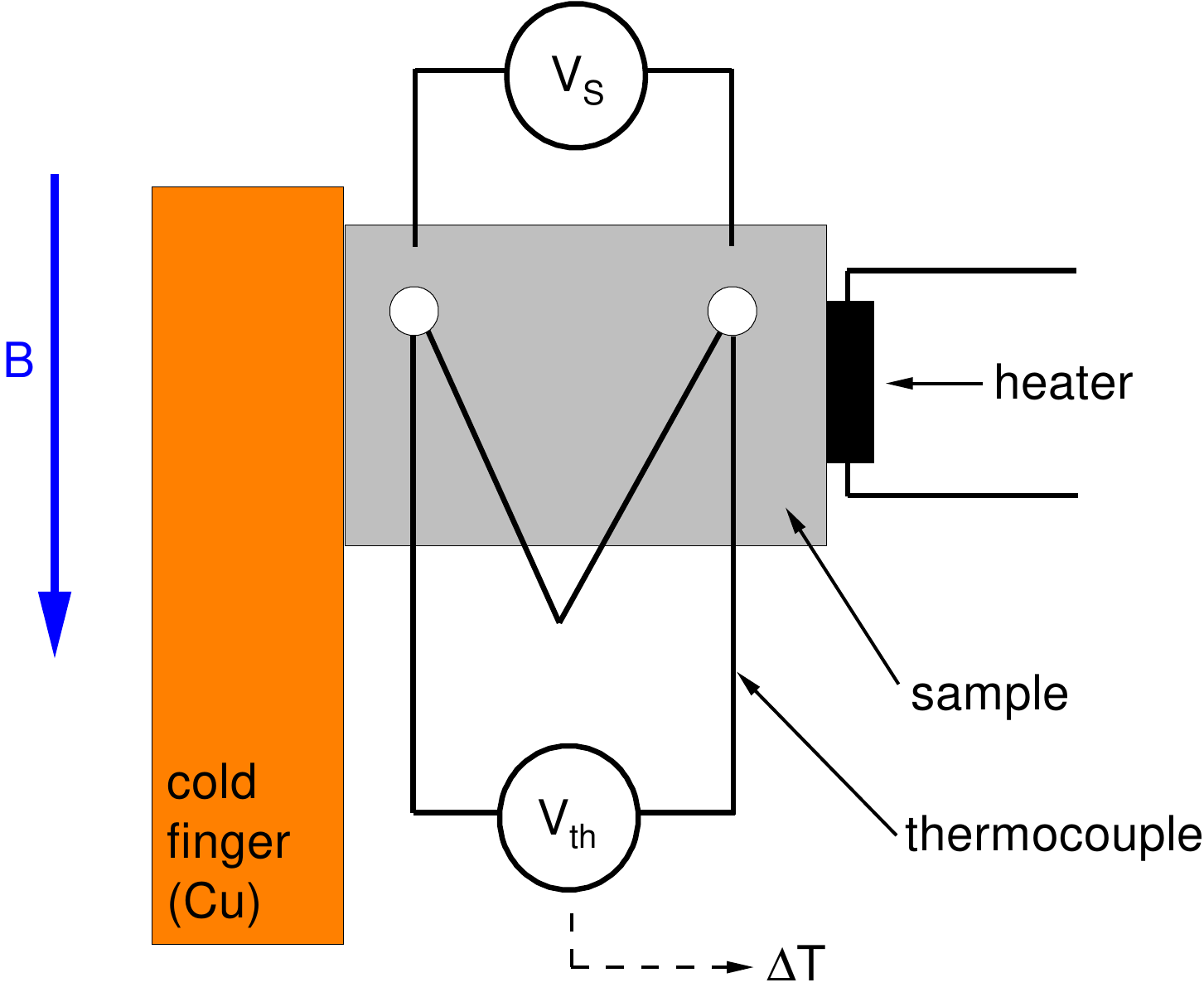}
		\caption{\textbf{Experimental setup:} A thermal gradient was applied along the trigonal axis and measured with a thermocouple. The Seebeck effect leads to a voltage $V_S$ parallel to the thermal gradient. The Seebeck coefficient is given by $S=-V_S/\Delta T$. The magnetic field was oriented parallel to the binary-bisectrix plane, i.e.\ perpendicular to the thermal gradient.}
		\label{fig:setup}
	\end{figure}
	
	The Seebeck coefficient $S_{zz}$ was measured with a homemade sample holder in a Quantum Design PPMS. As shown in Fig.~\ref{fig:setup}, a thermal gradient was applied along the trigonal axis using a RuO$_2$ heater and a cold finger made out of copper. The resulting temperature difference $\Delta T$ was determined with a type E thermocouple. The voltage contacts, which were made out of silver paste, were connected to the sample holder with manganin wires. Therefore, the resulting Seebeck coefficient $S=-V_S/\Delta T$ was corrected for the contribution of manganin according to Ref.~\cite{Rathnayaka1985}. The bismuth sample of purity 99.999\% (5N) with a length of 4~mm and a cross sectional area of 6.9~mm$^2$ was obtained commercially through MaTecK GmbH. This single crystal is of very high quality, which is reflected by a residual resistance ratio $R(300~\mathrm{K})/R(2~\mathrm{K})=576$, corresponding to an average mobility of $<\mu_e + \mu_h> =9.8\cdot10^7~\mathrm{cm^2\,V^{-1}\,s^{-1}}$ at very low temperatures. For angle dependent measurements, the rotator option of the PPMS was used. The sample was rotated such that the magnetic field was always lying in the binary-bisectrix plane (i.e.\ perpendicular to the thermal gradient) and the angle $\Theta$ is defined such that $\Theta = 0\degree$ for $B\parallel \mathrm{binary}$.

	\section{Results}\label{sec:results}
	\subsection{Zero-field Seebeck effect}
	
	\begin{figure*}
		\centering
		\includegraphics[width=1\linewidth]{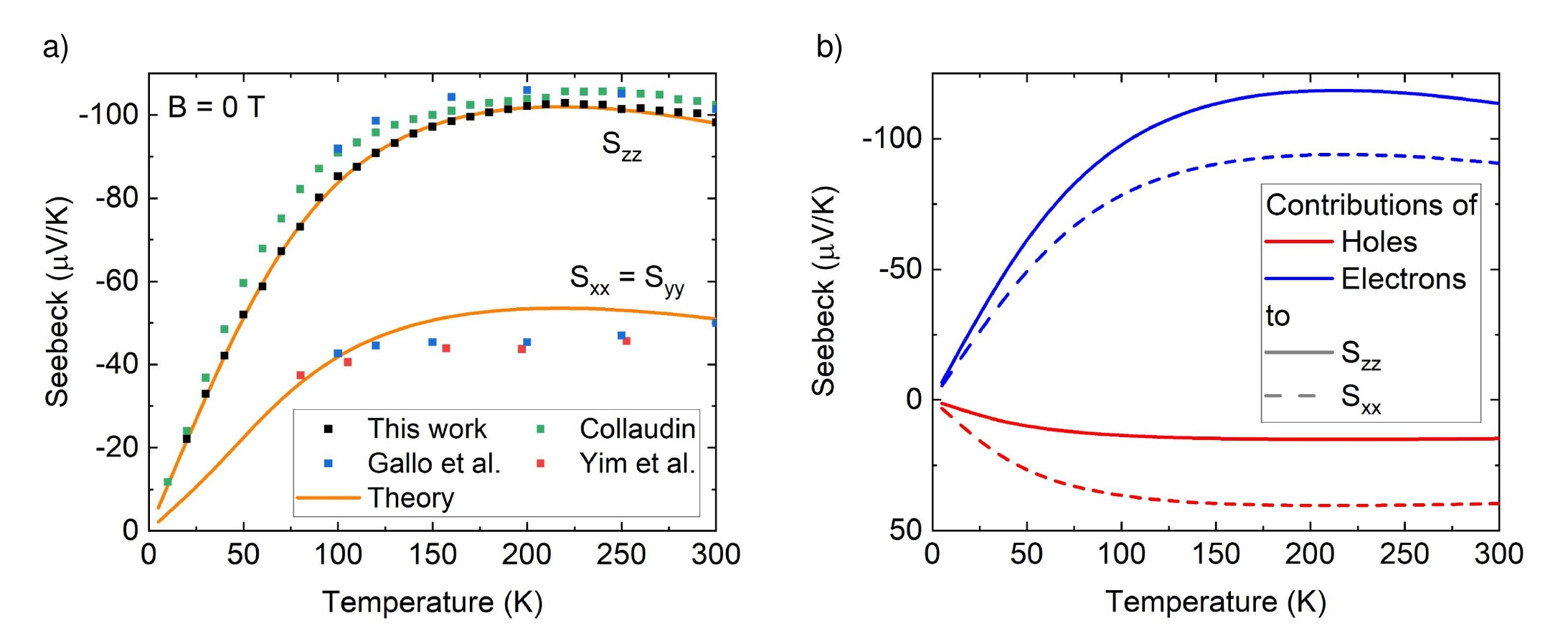}
		\caption{\textbf{Zero-field Seebeck coefficient vs.\ temperature:} a) Seebeck coefficient parallel to the trigonal axis ($S_{zz}$) and in the binary-bisectrix plane as a function of temperature at zero magnetic field. At low temperatures, the amplitude of the Seebeck effect is increasing linearly with temperature. Starting from $\approx 80$ K, it saturates to a very large value of $S_{zz}\approx -100\ \mu\mathrm{V\,K}^{-1}$. The theoretical curve is in very good agreement with the experimental data. Additional data is taken from Gallo et al.~\cite{Gallo1963}, Yim et al.~\cite{Yim1972} and Collaudin~\cite{Collaudin2014}. b) Contributions of holes and electrons to the zero-field thermoelectricity. It is dominated by the electrons due to their higher mobility.}
		\label{fig:SvsT}
	\end{figure*}
	
	The Seebeck coefficient $S_{zz}$ of bismuth is depicted in Fig.~\ref{fig:SvsT}a as a function of temperature $T$. It is negative, at $T < 80\ \mathrm{K}$ almost proportional to the temperature and shows a plateau-like behavior upon further heating. In this temperature range, $S_{zz}$ amounts to approximately $-100\ \mu\mathrm{V\,K^{-1}}$, which is a very large absolute value for a conducting material.
	
	How can these striking features of the zero-field Seebeck effect be understood? Firstly, the contributions of the different pockets have to be separated. The three electron pockets can be treated together, because regarding $S_{zz}$, they are equivalent when no magnetic field is applied. This is done theoretically in Fig.~\ref{fig:SvsT}b. As can be seen there, $S_{zz}$ is dominated by the electrons, which results in the negative sign of the Seebeck coefficient. The reason behind this is the mobility, which is much higher for the electrons than for the holes. Accordingly, the smaller difference of the mobilities in the binary-bisectrix plane leads to a less pronounced domination of the electrons and therefore to a lower absolute value of $S_{xx}$. Secondly, a qualitative understanding of the temperature dependence and the order of magnitude of the Seebeck coefficient can be reached by considering the Mott formula
	\begin{equation}\label{eq:Mott}
		S_{zz}^p = \frac{\pi^2k_B^2}{3e}T\left.\frac{\partial \ln(\sigma_{zz}^p)}{\partial \epsilon}\right|_{\epsilon=\epsilon_F} \propto \frac{T}{T_F^p}
	\end{equation}
	even though the condition $k_B T\ll \epsilon_F$ is not fulfilled here~\cite{Behnia2015}. The low Fermi temperatures $T_F^p$ of bismuth lead to its large Seebeck effect. Moreover, from the change of the band structure with temperature shown in Fig.~\ref{fig:theory}b follows that $T/T_F^p$ is almost constant between 100~K and 300~K for both electrons and holes. This explains why $S(T)$ is almost flat in this temperature range.

	\subsection{Magneto-Seebeck effect}
	\subsubsection{General behavior, Umkehr effect}
	
	\begin{figure*}
		\centering
		\includegraphics[width=1\linewidth]{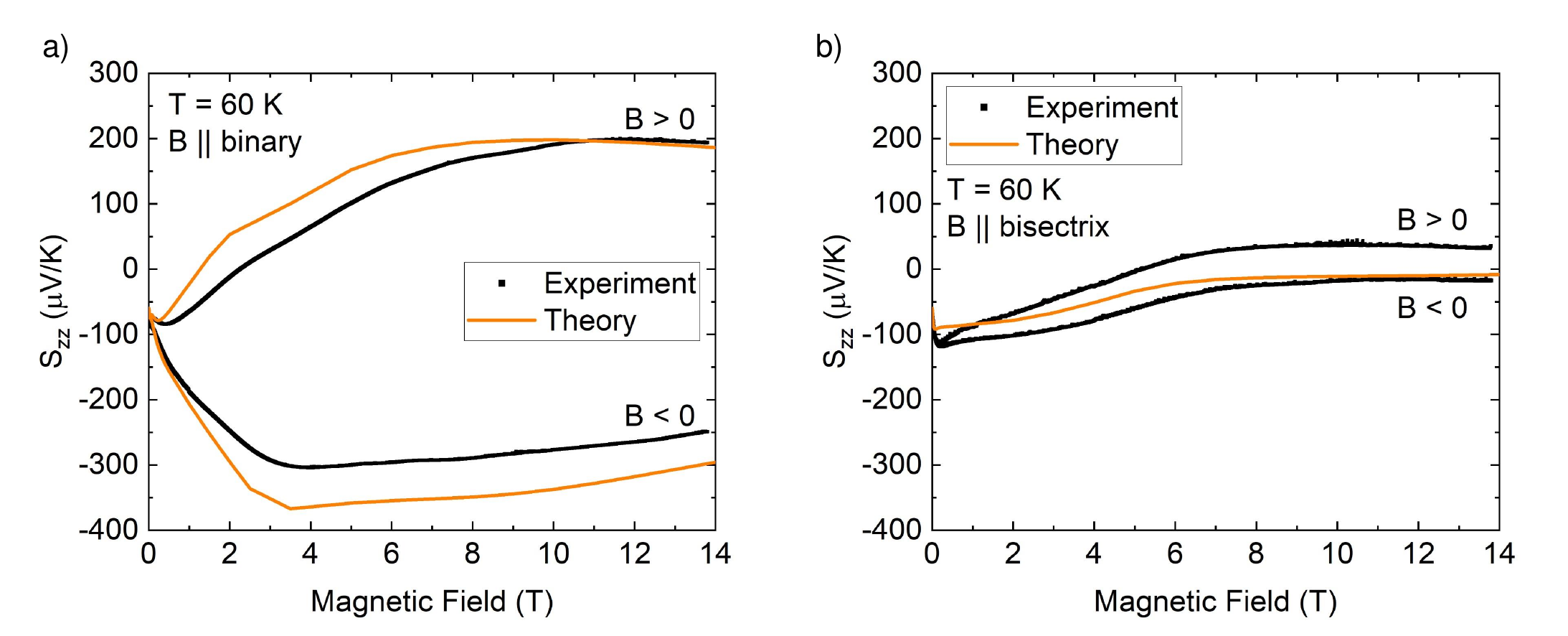}
		\caption{\textbf{Umkehr effect:} a) Seebeck coefficient vs.\ absolute magnetic field for $B\parallel\mathrm{binary}$ at $T=60$ K. A huge Umkehr effect can be observed, i.e.\ the Seebeck coefficient is not the same for positive and negative fields. b) As expected from symmetry considerations, the Umkehr effect is absent for $B\parallel\mathrm{bisectrix}$.}
		\label{fig:umkehr}
	\end{figure*}
	
	Fig.~\ref{fig:umkehr}a shows the Seebeck coefficient $S_{zz}$ at $T=60$~K as a function of the magnetic field applied parallel to a binary axis. The magnetic field strongly influences the Seebeck effect, e.g.\ a field of $B=-3$~T leads to a fivefold increase of $S_{zz}$. Furthermore, there is a large difference between positive and negative magnetic fields. At $B=+10$~T, the Seebeck coefficient amounts to $+191\ \mu\mathrm{V\,K^{-1}}$, whereas at $B=-10$~T, its value is $-277\ \mu\mathrm{V\,K^{-1}}$. 
	
	At this point one could argue that the difference between positive and negative fields is due to a misalignment of the voltage contacts, which would then lead to a contamination of the signal by the extremely large Nernst effect~\cite{Korenblit1969,Mangez1976,Behnia2007}. However, Fig.~\ref{fig:umkehr}b indicates that this is not the case. When the magnetic field is applied along a bisectrix axis, the difference between $S_{zz} (B)$ and $S_{zz} (-B)$ is much smaller than for $B\parallel\mathrm{binary}$.
	
	\begin{figure*}
		\centering
		\includegraphics[width=1\linewidth]{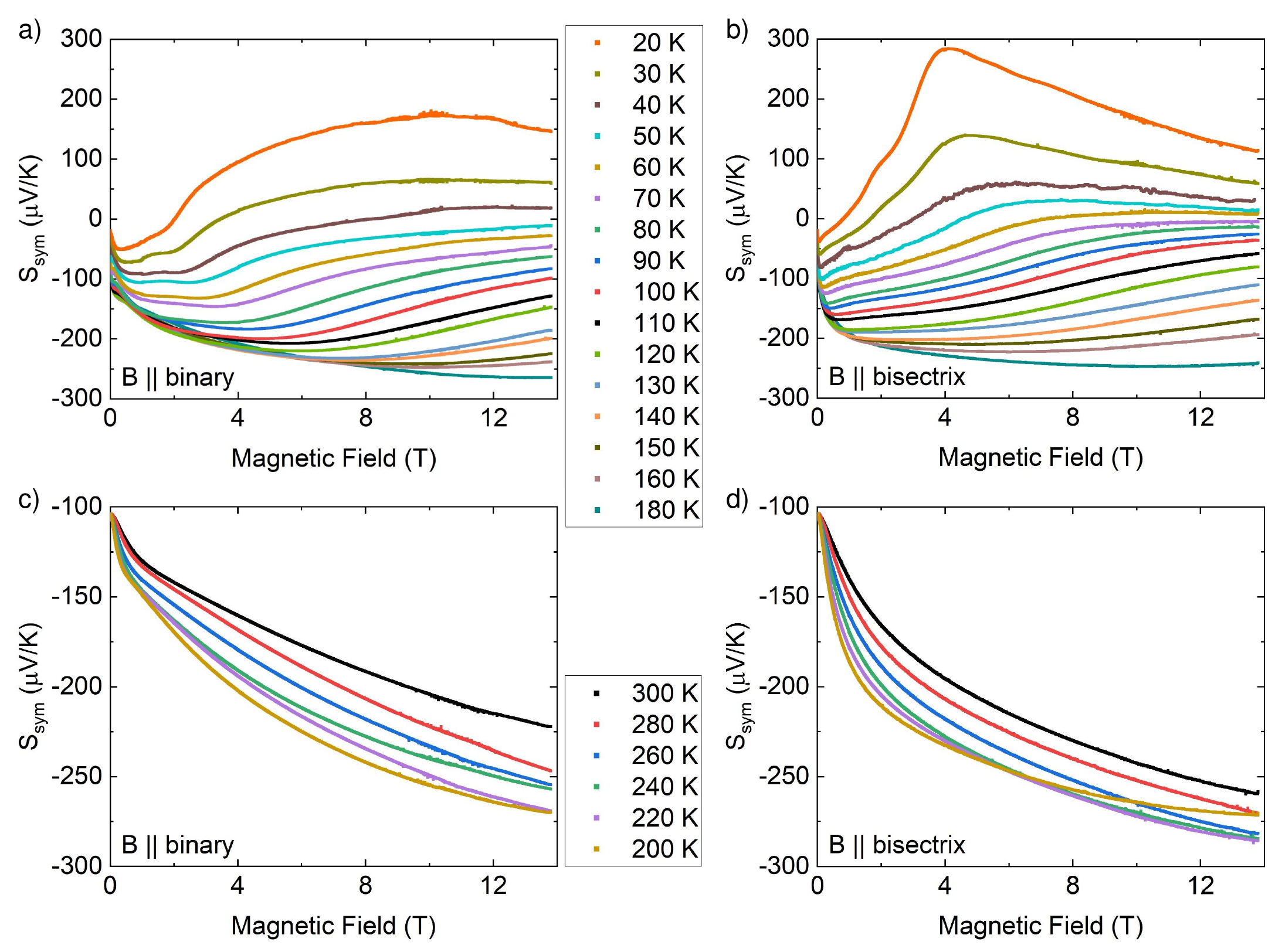}
		\caption{\textbf{Symmetrized Seebeck coefficient vs.\ magnetic field:} Symmetrized Seebeck coefficient $S_{sym}$ as a function of the magnetic field for $20~\mathrm{K} \leq T \leq 180~\mathrm{K}$ (upper panels) and $200~\mathrm{K} \leq T \leq 300~\mathrm{K}$ (lower panels). Measurements for $B\parallel\mathrm{binary}$ and $B\parallel\mathrm{bisectrix}$ are shown on the left and on the right, respectively.}
		\label{fig:SvsB_all}
	\end{figure*}
	
	As explained in section~\ref{sec:theory:general}, the Seebeck effect can show an Umkehr effect if the crystal symmetry is sufficiently low. This is possible for $B\parallel\mathrm{binary}$, whereas for $B\parallel\mathrm{bisectrix}$, the Umkehr effect is forbidden due to the crystal symmetry of bismuth~\cite{Wolfe1963,Akgoz1975}. The fact that the measured difference between $S_{zz} (B)$ and $S_{zz} (-B)$ is much smaller for $B\parallel\mathrm{bisectrix}$ than for $B\parallel\mathrm{binary}$ therefore strongly suggests that this unexpected behavior can be explained by the Umkehr effect. Moreover, the good theoretical description of the observed Umkehr effect also indicates that this effect is real. In this case, the difference between $S_{zz} (B)$ and $S_{zz} (-B)$ is given by the part of the off-diagonal contribution to equation~\eqref{eq:S_contributions}, which is odd in $B$. Hence, it becomes obvious from Fig.~\ref{fig:umkehr}a that the transversal contribution to the magneto-Seebeck effect is of great importance, at least for $B\parallel\mathrm{binary}$.
	
	For the small difference between the two experimental curves in Fig.~\ref{fig:umkehr}b, there are three conceivable reasons: a slight misalignment of the contacts, a small misalignment of the magnetic field and the bisectrix axis or an intrinsic Umkehr effect due to lifting of the twofold symmetry by magnetostriction~\cite{Michenaud2004}. In the following, we will only show data on and discuss the symmetrized Seebeck coefficient
	\begin{equation}
		S_{sym}(B) = \frac{S_{zz}(B)+S_{zz}(-B)}{2}
	\end{equation}
	in order to facilitate the comparison of the two field directions and to minimize the effect of a potential small misalignment of the voltage contacts.
	
	\subsubsection{Symmetrized Seebeck coefficient}
	
	Fig.~\ref{fig:SvsB_all} shows the full data set on $S_{sym}(B)$ for $B<13.8$~T oriented along the binary and bisectrix directions at $20\ \mathrm{K}\leq T \leq 300\ \mathrm{K}$. At temperatures above 200~K (panels c and~d), the symmetrized Seebeck coefficient gets monotonically more negative with increasing magnetic field. Moreover, the absolute value increases when lowering the temperature. Below 200~K (panels a and~b), on the other hand, lower temperatures lead to higher values of $S_{sym}$ and $S_{sym}$ is no longer a monotonic function of the magnetic field. As we will see below, this non-monotonic behavior is a consequence of Landau quantization. In general, the field dependence of the symmetrized Seebeck coefficient is more pronounced when the field is aligned with a bisectrix axis as when it is parallel to a binary axis. Note that this statement only holds true for $S_{sym}$, but not for the actual Seebeck coefficient $S_{zz}$ including the Umkehr effect (cf. Fig.~\ref{fig:umkehr}).

	\begin{figure*}
		\centering
		\includegraphics[width=1\linewidth]{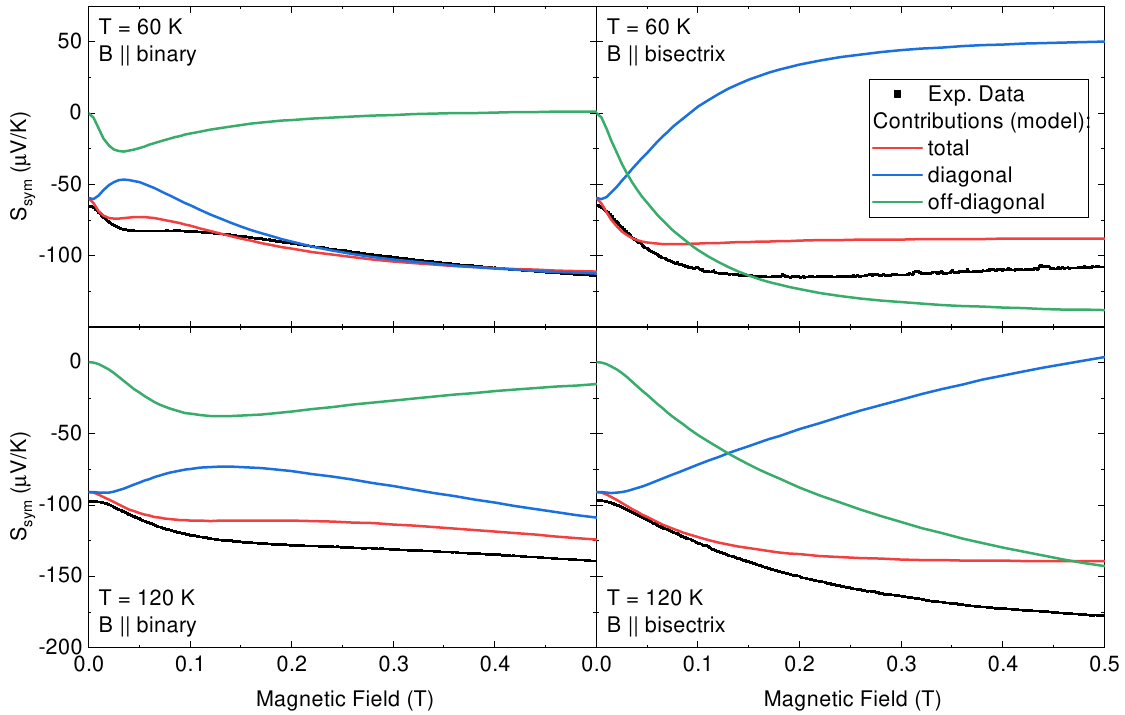}
		\caption{\textbf{Seebeck effect at low magnetic fields:} Symmetrized Seebeck coefficient as a function of the magnetic field for $B<0.5$ T. Experimental data is depicted in black. Coloured lines show the theoretical prediction (red) as well as its two components: The ordinary longitudinal component (blue) and the transversal contribution (green). The latter is important for both field orientations and even dominating for $B\parallel\mathrm{bisectrix}$.}
		\label{fig:SvsB_Nernst}
	\end{figure*}

	\subsubsection{Longitudinal and transversal contribution}
	
	The symmetrized Seebeck coefficient at low magnetic fields is depicted in Fig.~\ref{fig:SvsB_Nernst} for both field orientations and two temperatures (60~K and 120~K). It exhibits a non-trivial behavior: $S_{sym}$ is flat in a very narrow field window around 0~T, then the absolute value increases and at slightly higher fields, it starts to flatten again. In the case of $B\parallel\mathrm{binary}$ at 60~K, there is even a plateau in between (around $B=60$~mT). Increasing the temperature leads to a less pronounced response to the magnetic field. All of these features are captured by the calculations. To reach a better understanding of the underlying physics, the diagonal and off-diagonal components of the theoretical result are also plotted in Fig.~\ref{fig:SvsB_Nernst} (cf.\ Eq.~\eqref{eq:S_contributions}). When the magnetic field is applied along a bisectrix axis, the transversal contribution is clearly dominating as it sets the sign of the slope and at higher fields also the sign of the Seebeck coefficient itself. For $B\parallel\mathrm{binary}$, the absolute value of the transversal contribution to $S_{sym}$ is lower than the one of the longitudinal contribution, but it sets its slope at very low fields and is needed to explain the plateau close to $B=60$~mT. Taking into account the relevance of the off-diagonal component for the Umkehr effect, it becomes clear that the off-diagonal component, which is commonly ascribed to the Nernst and Hall effects, is indispensable to explain the magneto-Seebeck effect for both field orientations.

	\subsubsection{Angular dependence}
	
	\begin{figure*}
		\centering
		\includegraphics[width=1\linewidth]{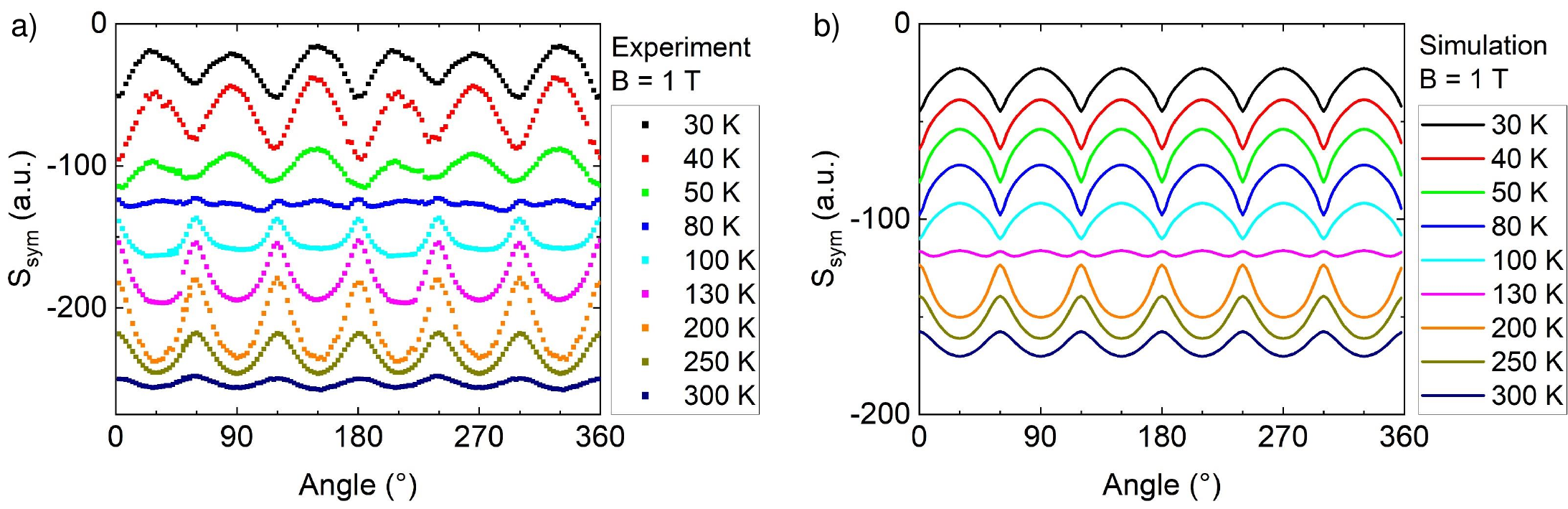}
		\caption{\textbf{Angle-dependent magneto-Seebeck effect:} a) Symmetrized Seebeck coefficient $S_{sym}$ along the trigonal axis vs.\ orientation of the magnetic field ($B=1\ \mathrm{T}$). Note that at high temperatures, $S_{sym}$ is maximal for $B\parallel \mathrm{binary}$ ($\Theta = 0\degree$) and minimal for $B\parallel \mathrm{bisectrix}$ ($\Theta = 30\degree$). This disagrees with the prediction of Ref.~\cite{Popescu2012}. b) Corresponding theoretical curves.} 
		\label{fig:SvsTheta}
	\end{figure*}
	
	Having understood the low field magneto-Seebeck effect in the cases when the magnetic field is oriented parallel to a main crystallographic axis, it is straightforward to compute the angular dependence of the Seebeck coefficient by choosing the tensor $\hat{B}$ in equations~\eqref{eq:sigma} and~\eqref{eq:alpha} accordingly. The result, which is depicted in Fig.~\ref{fig:SvsTheta}b, reproduces well the experimental data displayed in Fig.~\ref{fig:SvsTheta}a. The symmetrized Seebeck coefficient shows sixfold symmetry as expected from the symmetry of the Fermi surface. For $B=1$~T, at low temperatures, it is minimal for $B\parallel\mathrm{binary}$. Upon heating, the maxima and minima get inverted at 80~K (theory:\ 130~K) and the angular dependence is most pronounced roughly around 200~K. Further heating reduces the difference between peaks and dips, but $S_{sym} (\Theta)$ still shows maxima for $B\parallel\mathrm{binary}$.

	\subsubsection{Landau quantization}\label{sec:results_Landau}
	
	\begin{figure*}
		\centering
		\includegraphics[width=1\linewidth]{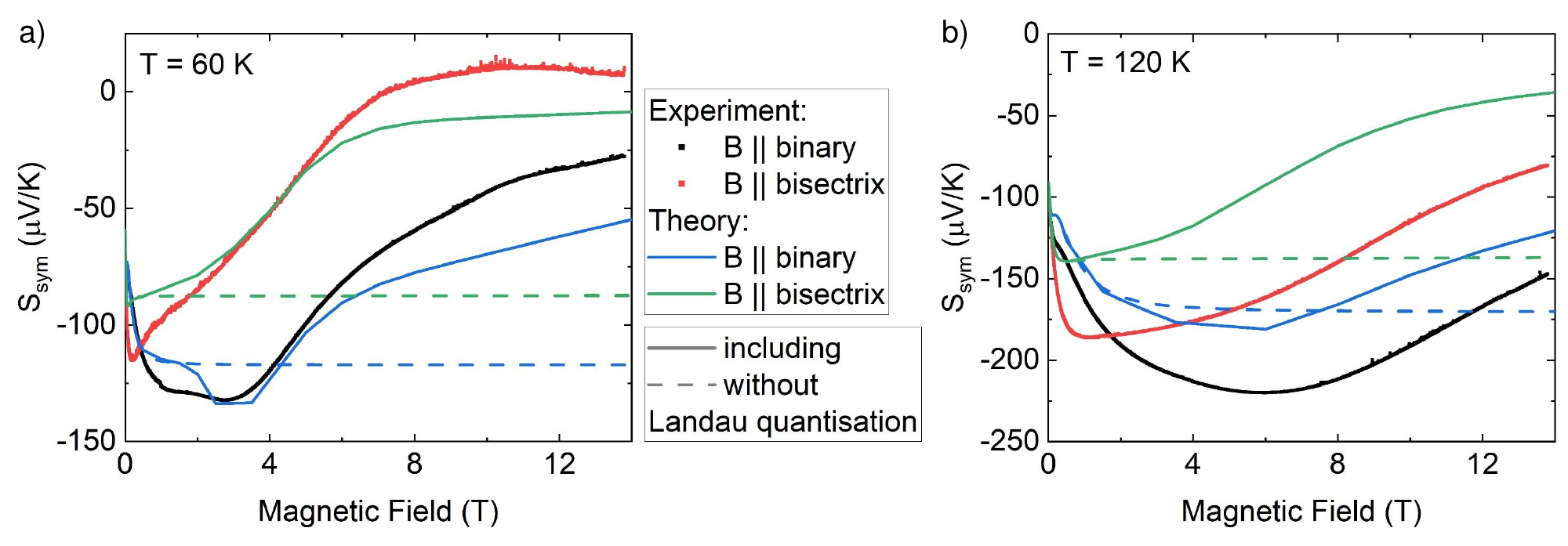}
		\caption{\textbf{Seebeck effect at high magnetic fields:} Symmetrized Seebeck coefficient as a function of the magnetic field for $B<13.8$ T. Experimental data is depicted in black ($B\parallel\mathrm{binary}$) and red ($B\parallel\mathrm{bisectrix}$). Blue and green lines show the theoretical prediction. Dashed lines correspond to the purely semiclassical theory without Landau quantization, whereas solid lines include it. The magneto-Seebeck effect of bismuth is clearly affected by Landau quantization at both 60~K and 120~K.}
		\label{fig:SvsB_Landau}
	\end{figure*}
	
	So far, we focused on the Seebeck effect in low magnetic fields up to 1~T. But what happens if the field strength is raised further? As shown in Fig.~\ref{fig:SvsB_Landau}, $S_{sym}$ approaches zero at high magnetic fields and even becomes slightly positive above 7.2~T for $B\parallel\mathrm{bisectrix}$ at 60~K. The minimum of the Seebeck coefficient is located at lower fields for $B\parallel\mathrm{bisectrix}$ than for $B\parallel\mathrm{binary}$ and shifts for both orientations to higher fields upon heating. Moreover, $S_{sym}$ is larger for $B\parallel\mathrm{bisectrix}$ than for $B\parallel\mathrm{binary}$ at high magnetic fields, whereas it is the other way round at low fields.
	
	The results of the purely semiclassical model are indicated by dashed lines in Fig.~\ref{fig:SvsB_Landau}. They are obviously not appropriate to describe the experimental data both at 60~K and 120~K as they quickly saturate at strongly negative values.
	
	\begin{figure*}
		\centering
		\includegraphics[width=0.86\linewidth]{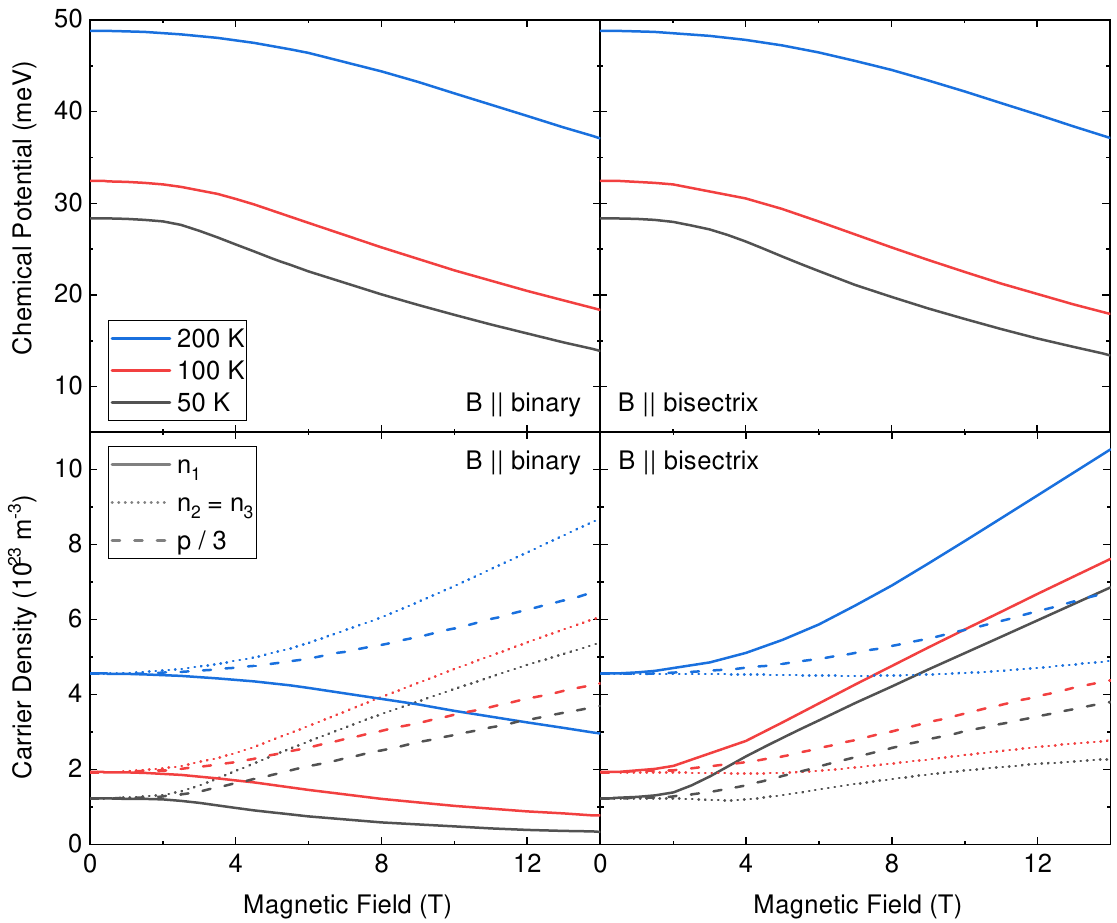}
		\caption{\textbf{Field-dependence of chemical potential and carrier density:} Theoretically determined values of the chemical potential (top) and the carrier density (bottom) as a function of the magnetic field for $B\parallel \mathrm{binary}$ (left) and $B\parallel \mathrm{bisectrix}$ (right) at several temperatures. Landau quantization changes the density of states. Therefore the chemical potential has to shift in order to keep the electron density $n=n_{1}+n_{2}+n_{3}$ equal to the hole density $p$. Note that only one third of the hole density is plotted in the lower panels.}
		\label{fig:mu_nvsB}
	\end{figure*}
	
	In contrast, if Landau quantization is taken into account, the theory qualitatively captures the features described above (see solid lines in Fig.~\ref{fig:SvsB_Landau}). The quantitative differences between model and theory are smaller at 60~K than at 120~K, but in both cases Landau quantization is essential to reproduce the experimentally observed non-monotonic behavior. This shows that Landau quantization significantly affects the Seebeck effect at both temperatures. At first glance, it is very surprising that a transport property is strongly influenced by Landau quantization at temperatures as high as 120~K, because generally it only plays a role at much lower temperatures. Let us see in the following why this is the case for the Seebeck effect in bismuth.
	
	Because of the extremely low carrier density and the strong anisotropy of the electron pockets in bismuth, the quantum limit of electrons is unusually small. This means that a magnetic field of only 1.3~T and 1.6~T along the bisectrix and binary direction, respectively, is sufficient to confine all electrons of at least one pocket to the lowest Landau level (see Eq.~\eqref{eq:dispersion_Landau} and Ref.~\cite{Zhu2011}). Further increasing the magnetic field above the quantum limit leads to an accumulation of electrons in the lowest Landau level. In order to keep charge compensation, the chemical potential decreases (see upper panels of Fig.~\ref{fig:mu_nvsB}). This change of about 15~meV (from 0~T to 14~T) affects the carrier density of all pockets and not only of the one which has reached the quantum limit (see lower panels of Fig.~\ref{fig:mu_nvsB}). The Mott formula~\eqref{eq:Mott} gives a qualitative account of the impact on the hole pocket's contribution to thermoelectricity: As $T_F$ increases at constant $T$, its Seebeck coefficient decreases.
	
\begin{figure}
		\centering
		\includegraphics[width=1\linewidth]{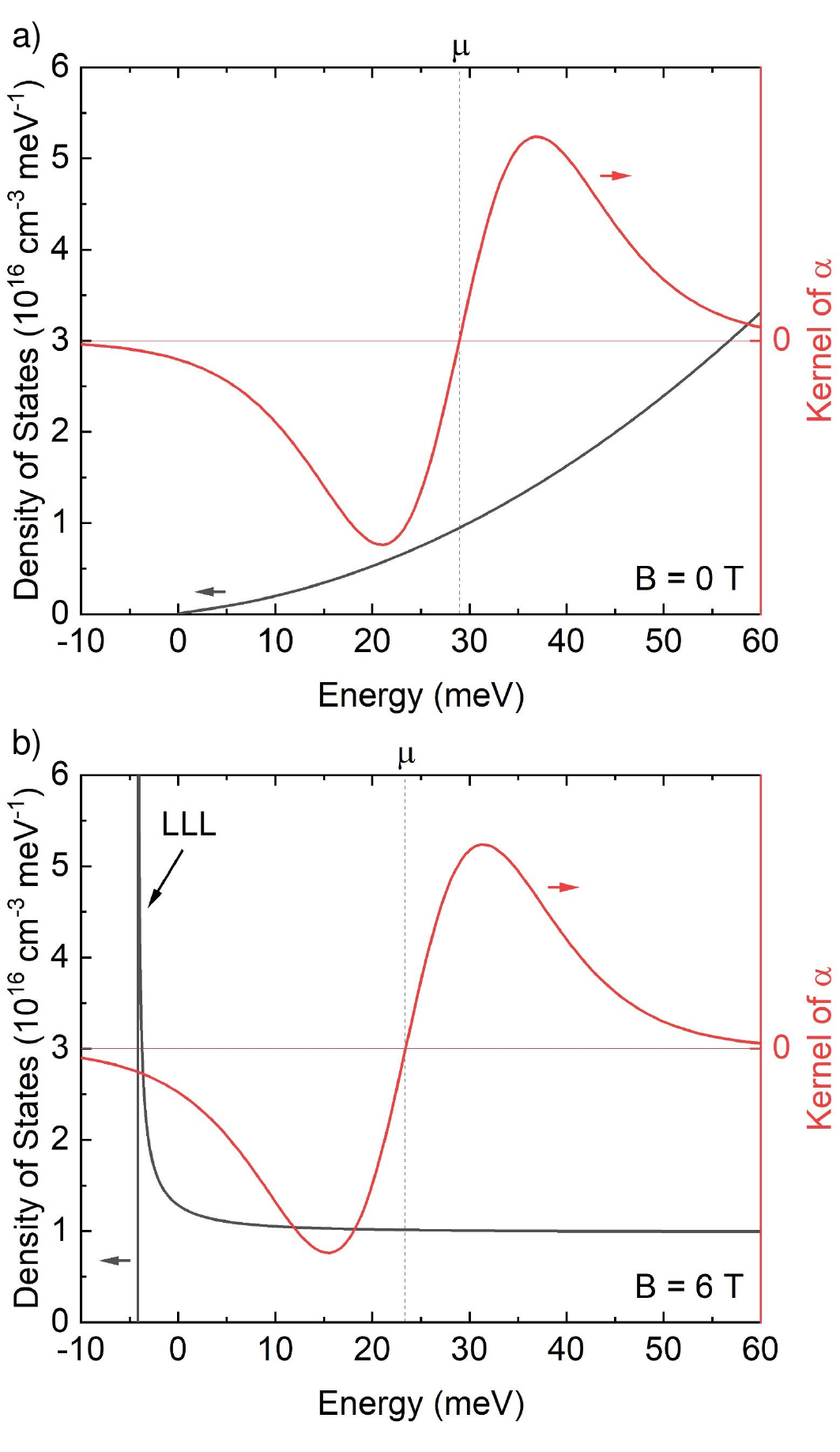}
		\caption{\textbf{Vanishing of the Seebeck effect in the quantum limit:} Density of states (black) in a) zero magnetic field and b) a field of 6~T along the bisectrix axis. The red curves depict the kernel $-\frac{(\epsilon - \mu)}{k_B T}\frac{\partial f^0}{\partial \epsilon}$ of the integral used to determine the thermoelectric conductivity~$\hat{\alpha}$ (cf.\ eqs.~\eqref{eq:alpha} and~\eqref{eq:alpha_Lax_Landau}) for $T=60$~K. Put simply, the Seebeck effect measures the integral of the product of this kernel and the density of states, i.e.\ the difference between the density of states above and below the chemical potential $\mu$ indicated by the broken line~\cite{Behnia2015}. In the quantum limit, the Seebeck effect vanishes upon increasing the magnetic field, because all Landau levels except the lowest Landau level (LLL) move to higher energies and the integral therefore approaches zero.}
		\label{fig:DOS}
	\end{figure}

	For the electrons in the quantum limit, another mechanism is crucial. In general, the Seebeck effect measures the difference between the density of states above and below the chemical potential~\cite{Behnia2015}. Roughly speaking, the thermoelectric counductivity~$\hat{\alpha}$ (and hence the Seebeck coefficient) is given by the integral over the kernel $-\frac{(\epsilon - \mu)}{k_B T}\frac{\partial f^0}{\partial \epsilon}$ times the density of states (see Fig.~\ref{fig:DOS} and eqs.~\eqref{eq:alpha} and~\eqref{eq:alpha_Lax_Landau}). Now, the density of states is fundamentally changed by Landau quantization: Electrons are moved from above the chemical potential to the lowest Landau level (far below the chemical potential). Therefore, in the vicinity of the chemical potential~$\mu$, the difference between $D(\epsilon)$ for $\epsilon > \mu$ and for $\epsilon < \mu$ essentially disappears and thus the contribution to the Seebeck effect of the respective pocket almost vanishes. This still holds true at $T=120$~K even though the term $\frac{\partial f^0}{\partial \epsilon}$ is thermally broadened, because the sign change of the term $\epsilon - \mu$ at the chemical potential $\mu$ and thus the shape of the curve depicting the kernel in Fig.~\ref{fig:DOS} is independent of temperature. Since the Seebeck effect is dominated by the electrons, the measured Seebeck coefficient also approaches zero.
	
	\section{Discussion}
	
	The experimental data on the zero-field Seebeck effect is in very good agreement with the values reported by Gallo et al.~\cite{Gallo1963} and Collaudin~\cite{Collaudin2014}. A small systematic error could be introduced by a misalignment of voltage and temperature contacts or by a slightly inhomogeneous heat flow.
	
	In a large part of the $(T,B,\Theta)$-space, the theory developed in section~\ref{sec:theory} is in good overall agreement with the experimental results presented in section~\ref{sec:results}. This is also true for the angular dependence of magnetoresistance and the Hall effect as shown in the supplement~\cite{Supplement}. For the zero-field Seebeck effect, the theoretically obtained values perfectly agree with the measurements of $S_{zz}$ and also match well the temperature dependence of $S_{xx}$ observed by Yim et al.~\cite{Yim1972} and Gallo et al.~\cite{Gallo1963}. In magnetic fields, the model works well above 50~K except for high magnetic fields at temperatures above 200~K (not shown). The agreement between experiment and theory is not as perfect as at zero field, but this would be rather mysterious given the choice we made in developing the model: We aimed at an understanding of the physical mechanisms via a model which is as simple as possible instead of perfectly reproducing experimental data by introducing a lot of adjustable parameters.
	
	One potential reason for the mismatch between theory and experiment at high magnetic fields above 200~K is the fact that it is not clear to what extent the Lax model (cf.\ Eq.~\eqref{eq:Lax}) is appropriate to approximate the band structure at high temperatures. Tight-binding calculations~\cite{Liu1995,Fuseya2015} suggest that the electron band bends within an energy window which could be relevant to the magneto-Seebeck effect at room temperature. We checked that taking Landau quantization out of the model does not solve this problem.
	
	At temperatures below 50~K, the main problem seems to be to get correct values for the Hall conductivities $\sigma_{ij}\ (i\neq j)$ which contribute to the Seebeck effect via the transversal component (cf.\ Eq.~\eqref{eq:S_contributions}). In general, it is quite difficult to predict the Hall effect of bismuth, because due to compensation the total Hall conductivity is the tiny difference of two very large values (for holes and electrons). For example, according to our model, at 60~K and 14~T, the total Hall conductivity amounts to less than $5\cdot 10^{-5}$ of the value for one carrier type. Hence, the predicted Hall conductivity is very susceptible to any changes to the model. Nevertheless, we achieve a good match between theory and the measured Hall effect above 50~K~\cite{Supplement}. At lower temperatures, one problem of our theory could be that we consider only electron-phonon scattering, but not electron-electron scattering. Furthermore, there is a recent report on a difference between bulk and surface conductance at low temperatures in bismuth~\cite{Kang2021}, which is out of the scope of the theory developed here. Lastly, phonon drag is important to the thermoelectricity of bismuth at very low temperatures~\cite{Mikhail1980,Uher1978,Issi1979}, but not taken into account here.
	
	As mentioned above, the Hall conductivity is very susceptible to changes to the model. This is particularly true for deviations from compensation. It was checked that tiny differences between the hole and the electron density affect the outcome of the calculations enormously. Therefore our results strongly suggest that bismuth is a perfectly compensated material. This implies that, contrary to what is sometimes assumed~\cite{Feng2021}, absence of compensation is not a prerequisite for a large Seebeck effect. 
	
	We note that Popescu and Woods~\cite{Popescu2012} calculated the angle-dependent magneto-Seebeck effect of bismuth for $100~\mathrm{K} < T < 300~\mathrm{K}$ and $B<2$~T. They predicted the Seebeck coefficient to be minimal for $B\parallel\mathrm{binary}$ and to increase with increasing magnetic field. Both these features are in contradiction with what was observed here. In addition, they failed to predict the Umkehr effect. From our point of view, the main reason why the model of Popescu and Woods conflicts with the experimental observations is the fact that they did not take into account the transversal contribution to the magneto-Seebeck effect.
	
	This transversal contribution was already implicitly included in the theory of Mikhail et al.~\cite{Mikhail1980}. The present work confirms their results for low magnetic fields. However, we explicitly point out the importance of the transversal contribution to the magneto-Seebeck effect: An applied thermal gradient gives rise to a transversal electric current. But due to the boundary conditions, this current cannot flow and instead, an electric field develops both in the transversal and the longitudinal direction. The one mentioned second, which is caused by the Hall resistivity, significantly impacts the measured Seebeck voltage and hence the value of the Seebeck coefficient. Since the transversal contribution increases the absolute value of the Seebeck coefficient, it is responsible for the strong increase of the thermoelectric figure of merit $ZT$ at low magnetic fields. This explanation probably also holds true for the doubling of $ZT$ under magnetic fields of a few hundred~mT in Bi-Sb alloys, which have the largest known thermoelectric figure of merit of any solid at cryogenic temperatures~\cite{Yim1972}.
	
	Furthermore, this work presents for the first time correct theoretical results on the angular dependence and the high-field behavior of the magneto-Seebeck effect of bismuth. The latter is achieved by including Landau quantization into the semiclassical transport theory, mainly by changing the dispersion relation and using the appropriate density of states. A priori, it was not clear if this procedure is allowed, but a posteriori, our results strongly suggest that it is a valid approach.
	
	Lastly, let us highlight the conceptual importance of the Umkehr effect. We observed a huge Umkehr effect for magnetic fields along the binary axis and were able to explain it theoretically. Although the Umkehr effect has been known in principle for decades~\cite{Akgoz1975,Wolfe1963,Michenaud1970}, it seems to us as this knowledge has got lost in parts of the community. Feng and Skinner recently wrote that because of Onsager reciprocity ``the value of the Seebeck coefficient is independent of the sign of the magnetic field''~\cite{Feng2021}. As we saw above, this is not the case here and in perfect agreement with Onsager reciprocity. This also means that extracting the value of the magneto-Seebeck effect by symmetrizing with respect to field inversion~\cite{Stockert2017,Gourgout2021} is only justified when there is no uncertainty about the alignment of the symmetry axes of the electron fluid and the underlying lattice.

	\section{Summary}
	
	We reported for the first time on a systematic study on the magneto-Seebeck effect of bismuth. In order to understand our experimental results, we developed a model based on semiclassical transport theory to which we added Landau quantization. In a large part of the $(T,B,\Theta)$-space, the calculations are in good agreement with experimental data on the zero-field Seebeck effect, the magneto-Seebeck effect, magnetoresistance and the Hall effect.
	
	We found that the large difference between the mobilities of electrons and holes as well as the temperature dependence of the band structure are essential to explain the zero-field Seebeck effect of bismuth. In magnetic fields, the tranverse contribution, which is composed of conductivity tensor entries that are commonly ascribed to the Nernst and Hall effects, plays an important role in setting the amplitude of the longitudinal Seebeck effect. It also gives rise to a large Umkehr effect, i.e.\ an odd-in-$B$ component of the magneto-Seebeck effect. At high magnetic fields, the Seebeck effect of bismuth is strongly affected by Landau quantization up to temperatures as high as 120~K.

	\section{Acknowledgements}
	
	F.S. thanks Liangcai Xu for his support on experimental issues related to thermocouples. This work was supported by the Agence Nationale de la Recherche (ANR-18-CE92-0020-01; ANR-19-CE30-0014-04), by Jeunes Equipes de l'Institut de Physique du Coll\`ege de France and by a grant attributed by the \^Ile-de-France regional council.
	
	\bibliography{Literature_MA}

	\clearpage
	\onecolumngrid
	\begin{center}
		\textbf{\large Supplementary Material}
	\end{center}
	\setcounter{equation}{0}
	\setcounter{figure}{0}
	\setcounter{table}{0}
	\setcounter{section}{0}
	\setcounter{page}{1}
	\makeatletter
	\renewcommand{\thesection}{S\arabic{section}}
	\renewcommand{\theequation}{S\arabic{equation}}
	\renewcommand{\thefigure}{S\arabic{figure}}
	
	\section{Derivation of the model of the transport properties of bismuth}
	
	\subsection{Conductivity tensor}\label{app:ch:conductivity}
	The scattering term of the Boltzmann equation is assumed to be
	\begin{equation}\label{app:scattering}
		\left.\frac{\partial f}{\partial t}\right|_{scattering} = \frac{\partial f^0}{\partial\epsilon}\mathbf{v}^\mathsf{T}\hat{\tau}^{-1}\boldsymbol{\psi}
	\end{equation}
	with $\boldsymbol{\psi}$ given by
	\begin{equation}\label{app:psi}
		f-f^0 = -\mathbf{v}^\mathsf{T}\boldsymbol{\psi}\frac{\partial f^0}{\partial\epsilon},
	\end{equation}
	where $f$, $f^0$ and $\mathbf{v}$ denote the distribution function, the Fermi-Dirac distribution and velocity, respectively~\cite{Mikhail1980}. The relaxation time $\hat{\tau}$ is a tensor. If $\hat{\tau} = \tau I_3$ (i.e.\ $\hat{\tau}$ is a scalar), this ansatz reduces to the standard expression for the relaxation-time approximation
	\begin{equation}
		\left.\frac{\partial f}{\partial t}\right|_{scattering} = -\frac{f-f^0}{\tau}.
	\end{equation}
	Using Eq.~\eqref{app:scattering}, the Boltzmann equation under presence of an electric field $\mathbf{E}$ and a magnetic field~$\mathbf{B}$ reads
	\begin{equation}
		-\frac{e}{\hbar}(\boldsymbol{\nabla}_{\!k} f)^\mathsf{T}\mathbf{E}-\frac{e}{\hbar}(\boldsymbol{\nabla}_{\!k} f)^\mathsf{T}(\mathbf{v}\times\mathbf{B}) = \frac{\partial f^0}{\partial\epsilon}\mathbf{v}^\mathsf{T}\hat{\tau}^{-1}\boldsymbol{\psi}.
	\end{equation}
	Replacing $f$ by $f^0$ in the first term (linearisation) and $f$ by $g = f - f^0$ in the second term (the magnetic field has only an influence on electrons which are already out of equilibrium) leads to
	\begin{equation}
		-\frac{e}{\hbar}(\boldsymbol{\nabla}_{\!k} f^0)^\mathsf{T}\mathbf{E}-\frac{e}{\hbar}(\mathbf{v}\times\mathbf{B})^\mathsf{T}(\boldsymbol{\nabla}_{\!k} g) = \frac{\partial f^0}{\partial\epsilon}\mathbf{v}^\mathsf{T}\hat{\tau}^{-1}\boldsymbol{\psi}.
	\end{equation}
	The velocity is given by $v_i = \hbar^{-1}  \partial \epsilon / \partial k_i$. Therefore one gets
	\begin{equation}
		-e\frac{\partial f^0}{\partial\epsilon}\mathbf{v}^\mathsf{T}\mathbf{E}-\frac{e}{\hbar^2}(\mathbf{v}\times\mathbf{B})^\mathsf{T}\widehat{\left(\frac{\partial^2 \epsilon}{\partial k_i \partial k_j}\right)}(\boldsymbol{\nabla}_{\!v} g) = \frac{\partial f^0}{\partial\epsilon}\mathbf{v}^\mathsf{T}\hat{\tau}^{-1}\boldsymbol{\psi}.
	\end{equation}
	The chain rule leads to
	\begin{equation}
		-e\frac{\partial f^0}{\partial\epsilon}\mathbf{v}^\mathsf{T}\mathbf{E}-\frac{e}{\hbar^2}(\mathbf{v}\times\mathbf{B})^\mathsf{T}\frac{\partial \epsilon}{\partial \gamma}\widehat{\left(\frac{\partial^2 \gamma}{\partial k_i \partial k_j}\right)}(\boldsymbol{\nabla}_{\!v} g) = \frac{\partial f^0}{\partial\epsilon}\mathbf{v}^\mathsf{T}\hat{\tau}^{-1}\boldsymbol{\psi}
	\end{equation}
	and
	\begin{equation}\label{app:eq11}
		-e\frac{\partial f^0}{\partial\epsilon}\mathbf{v}^\mathsf{T}\mathbf{E}-e(\mathbf{v}\times\mathbf{B})^\mathsf{T}\frac{1}{\gamma'}\hat{m}_{be}^{-1}(\boldsymbol{\nabla}_{\!v} g) = \frac{\partial f^0}{\partial\epsilon}\mathbf{v}^\mathsf{T}\hat{\tau}^{-1}\boldsymbol{\psi}.
	\end{equation}
	Using the antisymmetric magnetic field tensor $\hat{B}$ instead of $\mathbf{B}$, one can write
	\begin{equation}
		(\mathbf{v}\times\mathbf{B})^\mathsf{T} = -(\mathbf{B}\times\mathbf{v})^\mathsf{T} = -(\hat{B}\mathbf{v})^\mathsf{T} = -\mathbf{v}^\mathsf{T}\hat{B}^\mathsf{T} = \mathbf{v}^\mathsf{T}\hat{B}.
	\end{equation}
	By inserting this in Eq.~\eqref{app:eq11} we get
	\begin{equation}\label{app:eq18}
		\mathbf{v}^\mathsf{T}\left(-e\frac{\partial f^0}{\partial\epsilon}\mathbf{E}-\frac{e}{\gamma'}\hat{B}\hat{m}_{be}^{-1}(\boldsymbol{\nabla}_{\!v} g)\right) = \mathbf{v}^\mathsf{T}\hat{\tau}^{-1}\boldsymbol{\psi}\frac{\partial f^0}{\partial\epsilon}.
	\end{equation}
	Obviously, one possible solution to Eq.~\eqref{app:eq18} is
	\begin{equation}
		-e\frac{\partial f^0}{\partial\epsilon}\mathbf{E}-\frac{e}{\gamma'}\hat{B}\hat{m}_{be}^{-1}(\boldsymbol{\nabla}_{\!v} g) = \hat{\tau}^{-1}\boldsymbol{\psi}\frac{\partial f^0}{\partial\epsilon}
	\end{equation}
	\begingroup\abovedisplayshortskip=-1em
	\begin{equation}
		\Rightarrow \hat{\tau}\left(-e\frac{\partial f^0}{\partial\epsilon}\mathbf{E}-\frac{e}{\gamma'}\hat{B}\hat{m}_{be}^{-1}(\boldsymbol{\nabla}_{\!v} g)\right) = \boldsymbol{\psi}\frac{\partial f^0}{\partial\epsilon}
	\end{equation}\endgroup
	Inserting this in Eq.~\eqref{app:psi} gives
	\begin{equation}
		g = \mathbf{v}^\mathsf{T}\hat{\tau}\left(e\frac{\partial f^0}{\partial\epsilon}\mathbf{E}+\frac{e}{\gamma'}\hat{B}\hat{m}_{be}^{-1}(\boldsymbol{\nabla}_{\!v} g)\right).
	\end{equation}
	With
	\begin{equation}\label{app:Gvec}
		\mathbf{G} = e\frac{\partial f^0}{\partial\epsilon}\mathbf{E}+\frac{e}{\gamma'}\hat{B}\hat{m}_{be}^{-1}(\boldsymbol{\nabla}_{\!v} g)
	\end{equation}
	one can write
	\begin{equation}\label{app:gshort}
		g = \mathbf{v}^\mathsf{T}\hat{\tau}\mathbf{G}.
	\end{equation}
	Since $\hat{\tau}$ and $\mathbf{G}$ do not depend on $\mathbf{v}$, we have
	\begin{equation}
		\boldsymbol{\nabla}_{\!v} g = \hat{\tau}\mathbf{G},
	\end{equation}
	which can be inserted into Eq.~\eqref{app:Gvec}:
	\begin{equation}
		\mathbf{G} = e\frac{\partial f^0}{\partial\epsilon}\mathbf{E}+\frac{e}{\gamma'}\hat{B}\hat{m}_{be}^{-1}\hat{\tau}\mathbf{G}
	\end{equation}
	\begin{equation}
		\Rightarrow \mathbf{G}-\frac{e}{\gamma'}\hat{B}\hat{m}_{be}^{-1}\hat{\tau}\mathbf{G} = e\frac{\partial f^0}{\partial\epsilon}\mathbf{E}
	\end{equation}
	\begin{equation}
		\Rightarrow \left(I_3-\frac{e}{\gamma'}\hat{B}\hat{m}_{be}^{-1}\hat{\tau}\right)\mathbf{G} = e\frac{\partial f^0}{\partial\epsilon}\mathbf{E}
	\end{equation}
	\begin{equation}\label{app:Gresult}
		\Rightarrow \mathbf{G} = \left(I_3-\frac{e}{\gamma'}\hat{B}\hat{m}_{be}^{-1}\hat{\tau}\right)^{-1} e\frac{\partial f^0}{\partial\epsilon}\mathbf{E}
	\end{equation}
	Finally, inserting Eq.~\eqref{app:Gresult} into Eq.~\eqref{app:gshort} yields
	\begin{equation}\label{app:gExplicit}
		g = \mathbf{v}^\mathsf{T}\hat{\tau}\left(I_3-\frac{e}{\gamma'}\hat{B}\hat{m}_{be}^{-1}\hat{\tau}\right)^{-1} e\frac{\partial f^0}{\partial\epsilon}\mathbf{E}.
	\end{equation}
	The electric current density can be calculated by
	\begin{equation}\label{app:jSimple}
		\mathbf{j} = -\frac{e}{4\pi^3}\int \mathbf{v}g\ \mathrm{d}^3\!k.
	\end{equation}
	Combining equations \eqref{app:gExplicit} and \eqref{app:jSimple} leads to
	\begin{equation}
		\mathbf{j} = -\frac{e}{4\pi^3}\int \mathbf{v}\mathbf{v}^\mathsf{T}\hat{\tau}\left(I_3-\frac{e}{\gamma'}\hat{B}\hat{m}_{be}^{-1}\hat{\tau}\right)^{-1} e\frac{\partial f^0}{\partial\epsilon}\mathbf{E}\ \mathrm{d}^3\!k.
	\end{equation}
	The integrand can be simplified by inserting an identity matrix:
	\begin{equation}
		\mathbf{j} = -\frac{e}{4\pi^3}\int \mathbf{v}\mathbf{v}^\mathsf{T}\hat{\tau}\left(\hat{m}_{be}^{-1}\hat{\tau}\frac{e}{\gamma'}\right)^{-1}\left(\hat{m}_{be}^{-1}\hat{\tau}\frac{e}{\gamma'}\right)\left(I_3-\frac{e}{\gamma'}\hat{B}\hat{m}_{be}^{-1}\hat{\tau}\right)^{-1} e\frac{\partial f^0}{\partial\epsilon}\mathbf{E}\ \mathrm{d}^3\!k
	\end{equation}
	\begin{equation}
		\Rightarrow \mathbf{j} = -\frac{e}{4\pi^3}\int \mathbf{v}\mathbf{v}^\mathsf{T}\hat{\tau}\hat{\tau}^{-1}\hat{m}_{be}\gamma'\left(\left(\left(I_3-\frac{e}{\gamma'}\hat{B}\hat{m}_{be}^{-1}\hat{\tau}\right)^{-1}\right)^{-1}\left(\hat{m}_{be}^{-1}\hat{\tau}\frac{e}{\gamma'}\right)^{-1}\right)^{-1}   \frac{\partial f^0}{\partial\epsilon}\mathbf{E}\ \mathrm{d}^3\!k
	\end{equation}
	\begingroup\abovedisplayshortskip=-1em
	\begin{equation}
		\Rightarrow \mathbf{j} = -\frac{e}{4\pi^3}\int \mathbf{v}\mathbf{v}^\mathsf{T}\hat{m}_{be}\gamma'\left(\left(I_3-\frac{e}{\gamma'}\hat{B}\hat{m}_{be}^{-1}\hat{\tau}\right)\hat{\tau}^{-1}\hat{m}_{be}\frac{\gamma'}{e}\right)^{-1} \frac{\partial f^0}{\partial\epsilon}\mathbf{E}\ \mathrm{d}^3\!k
	\end{equation}\endgroup
	\begin{equation}
		\Rightarrow \mathbf{j} = -\frac{e}{4\pi^3}\int \widehat{\left(\frac{\partial\epsilon}{\partial k_i}\frac{\partial\epsilon}{\partial k_j}\right)}\widehat{\left(\frac{\partial^2\gamma}{\partial k_i \partial k_j}\right)}^{-1}\gamma'\left(\left(\frac{e}{\gamma'}\hat{m}_{be}^{-1}\hat{\tau}\right)^{-1}-\hat{B}\right)^{-1}  \frac{\partial f^0}{\partial\epsilon}\mathbf{E}\ \mathrm{d}^3\!k 
	\end{equation}
	\begin{equation}
		\Rightarrow \mathbf{j} = -\frac{e}{4\pi^3}\int \left(\frac{\partial \epsilon}{\partial \gamma}\right)^2 \widehat{\left(\frac{\partial\gamma}{\partial k_i}\frac{\partial\gamma}{\partial k_j}\right)}\widehat{\left(\frac{\partial^2\gamma}{\partial k_i \partial k_j}\right)}^{-1}\gamma'\left(\left(\frac{e}{\gamma'}\hat{m}_{be}^{-1}\hat{\tau}\right)^{-1}-\hat{B}\right)^{-1} \frac{\partial f^0}{\partial\epsilon}\mathbf{E}\ \mathrm{d}^3\!k 
	\end{equation}
	\begin{equation}
		\Rightarrow \mathbf{j} = -\frac{e}{4\pi^3}\int \frac{1}{\gamma'} \widehat{\left(\frac{\partial\gamma}{\partial k_i}\frac{\partial\gamma}{\partial k_j}\right)}\widehat{\left(\frac{\partial^2\gamma}{\partial k_i \partial k_j}\right)}^{-1}\left(\left(\frac{e}{\gamma'}\hat{m}_{be}^{-1}\hat{\tau}\right)^{-1}-\hat{B}\right)^{-1}  \frac{\partial f^0}{\partial\epsilon}\mathbf{E}\ \mathrm{d}^3\!k
	\end{equation}
	In the following, we assume for simplicity that the main axes of the ellipsoid representing the Fermi surface are parallel to the coordinate axes (i.e. $\hat{m}_{be}^{-1}$ is diagonal). The derivatives of $\gamma$ can be computed using Eq.~\eqref{eq:Lax_gamma}.
	\begin{equation}
		\begin{split}
			\mathbf{j} = -\frac{e}{4\pi^3}\int & \frac{1}{\gamma'} \hbar^4
			\begin{pmatrix}
				(k_x m_{be_{xx}}^{-1})^2 & k_x m_{be_{xx}}^{-1} k_y m_{be_{yy}}^{-1} & k_x m_{be_{xx}}^{-1} k_z m_{be_{zz}}^{-1}\\
				k_x m_{be_{xx}}^{-1} k_y m_{be_{yy}}^{-1} & (k_y m_{be_{yy}}^{-1})^2 & k_y m_{be_{yy}}^{-1} k_z m_{be_{zz}}^{-1}\\
				k_x m_{be_{xx}}^{-1} k_z m_{be_{zz}}^{-1} & k_y m_{be_{yy}}^{-1} k_z m_{be_{zz}}^{-1} & (k_z m_{be_{zz}}^{-1})^2\\
			\end{pmatrix}\\
			& \cdot \hbar^{-2}
			\begin{pmatrix}
				m_{be_{xx}} & 0 & 0\\
				0 & m_{be_{yy}} & 0\\
				0 & 0 & m_{be_{zz}}\\
			\end{pmatrix}
			\left(\left(\frac{e}{\gamma'}\hat{m}_{be}^{-1}\hat{\tau}\right)^{-1}-\hat{B}\right)^{-1} \frac{\partial f^0}{\partial\epsilon}\mathbf{E}\ \mathrm{d}^3\!k
		\end{split}
	\end{equation}
	All terms of the integrand except for the first matrix are symmetric functions of $k_i$. Since the off-diagonal elements of the first matrix are anti-symmetric in $k_i$, they give rise to integrands, which are anti-symmetric functions of $k_i$. As the integration is performed over the whole k-space, the integral of anti-symmetric functions of $k_i$ is zero. This is why the off-diagonal elements can be set to zero.
	\begin{equation}
		\begin{split}
			\mathbf{j} = -\frac{e}{4\pi^3}\int & \frac{1}{\gamma'} \hbar^2
			\begin{pmatrix}
				(k_x m_{be_{xx}}^{-1})^2 & 0 & 0\\
				0 & (k_y m_{be_{yy}}^{-1})^2 & 0\\
				0 & 0 & (k_z m_{be_{zz}}^{-1})^2\\
			\end{pmatrix}\\
			&\cdot
			\begin{pmatrix}
				m_{be_{xx}} & 0 & 0\\
				0 & m_{be_{yy}} & 0\\
				0 & 0 & m_{be_{zz}}\\
			\end{pmatrix}
			\left(\left(\frac{e}{\gamma'}\hat{m}_{be}^{-1}\hat{\tau}\right)^{-1}-\hat{B}\right)^{-1} \frac{\partial f^0}{\partial\epsilon}\mathbf{E}\ \mathrm{d}^3\!k
		\end{split}
	\end{equation}
	\begin{equation}
		\Rightarrow\mathbf{j} = -\frac{e}{4\pi^3}\int \frac{\hbar^2}{\gamma'} 
		\begin{pmatrix}
			k_x^2 m_{be_{xx}}^{-1} & 0 & 0\\
			0 & k_y^2 m_{be_{yy}}^{-1} & 0\\
			0 & 0 & k_z^2 m_{be_{zz}}^{-1}\\
		\end{pmatrix}
		\left(\left(\frac{e}{\gamma'}\hat{m}_{be}^{-1}\hat{\tau}\right)^{-1}-\hat{B}\right)^{-1}  \frac{\partial f^0}{\partial\epsilon}\mathbf{E}\ \mathrm{d}^3\!k \label{app:eq:start_Landau}
	\end{equation}
	For switching from integration over wavenumbers to integration over energy, one needs the density of states $D(\epsilon)$, which is given in Eq.~\eqref{eq:DOS}. Furthermore, $\hbar^2 k_i^2 m_{be_{ii}}^{-1}$ can be replaced by $2\gamma/3$ in this step.
	\begin{equation}
		\mathbf{j} = -e\int \frac{1}{\gamma'} 
		\begin{pmatrix}
			2\gamma/3 & 0 & 0\\
			0 & 2\gamma/3 & 0\\
			0 & 0 & 2\gamma/3\\
		\end{pmatrix}
		\left(\left(\frac{e}{\gamma'}\hat{m}_{be}^{-1}\hat{\tau}\right)^{-1}-\hat{B}\right)^{-1} \frac{\partial f^0}{\partial\epsilon}\mathbf{E} D(\epsilon)\ \mathrm{d}\epsilon
	\end{equation}
	\begin{equation}
		\Rightarrow \mathbf{j} = -\sqrt{\frac{2}{\det \hat{m}_{be}^{-1}}} \frac{2e}{3\pi^2\hbar^3} \int \gamma^{3/2}
		\left(\left(\frac{e}{\gamma'}\hat{m}_{be}^{-1}\hat{\tau}\right)^{-1}-\hat{B}\right)^{-1} \frac{\partial f^0}{\partial\epsilon}\ \mathrm{d}\epsilon\ \mathbf{E}
	\end{equation}
	Therefore, the conductivity tensor $\hat{\sigma}$, which relates current density and electric field by $\mathbf{j} = \hat{\sigma} \mathbf{E}$ is given as follows:
	\begin{equation}\label{app:eq:sigma}
		{\hat{\sigma} = \sum_{\mathrm{pockets}} -\sqrt{\frac{2}{\det \hat{m}_{be}^{-1}}} \frac{2e}{3\pi^2\hbar^3} \int \gamma^{3/2}
			\left(\left(\frac{e}{\gamma'}\hat{m}_{be}^{-1}\hat{\tau}\right)^{-1}-\hat{B}\right)^{-1} \frac{\partial f^0}{\partial\epsilon}\ \mathrm{d}\epsilon}
	\end{equation}
	As there is more than one carrier pocket, the contributions of all pockets have to be summed up. Except for the sign of $\hat{B}$, this result is equivalent to equation 8a in Ref.~\cite{Mikhail1980}.
	
	\subsection{Thermoelectricity tensor}
	Using the same ansatz as in section~\ref{app:ch:conductivity}, the Boltzmann equation under presence of a thermal gradient $\boldsymbol{\nabla}T$ and a magnetic field $\mathbf{B}$ reads
	\begin{equation}
		-\frac{\partial f}{\partial T}(\boldsymbol{\nabla} T)^\mathsf{T}\mathbf{v}-\frac{e}{\hbar}(\boldsymbol{\nabla}_{\!k} f)^\mathsf{T}(\mathbf{v}\times\mathbf{B}) = \frac{\partial f^0}{\partial\epsilon}\mathbf{v}^\mathsf{T}\hat{\tau}^{-1}\boldsymbol{\psi}.
	\end{equation}
	Linearisation and rewriting the first term using the chemical potential $\mu$ leads to
	\begin{equation}
		-\frac{\epsilon-\mu}{T}\frac{\partial f^0}{\partial \epsilon}(\boldsymbol{\nabla} T)^\mathsf{T}\mathbf{v}-\frac{e}{\hbar}(\boldsymbol{\nabla}_{\!k} f)^\mathsf{T}(\mathbf{v}\times\mathbf{B}) = \frac{\partial f^0}{\partial\epsilon}\mathbf{v}^\mathsf{T}\hat{\tau}^{-1}\boldsymbol{\psi}.
	\end{equation}
	Starting from this equation, the thermoelectricity tensor $\hat{\alpha}$, which relates electric current density and thermal gradient by $\mathbf{j} = -\hat{\alpha}\boldsymbol{\nabla}T$ can be derived performing the same steps as in section~\ref{app:ch:conductivity}. This derivation gives:
	\begin{equation}\label{app:eq:alpha}
		{\hat{\alpha} = \sum_{\mathrm{pockets}} -\sqrt{\frac{2}{\det \hat{m}_{be}^{-1}}} \frac{2}{3\pi^2\hbar^3} \int \frac{\epsilon-\mu}{T} \gamma^{3/2}
			\left(\left(\frac{e}{\gamma'}\hat{m}_{be}^{-1}\hat{\tau}\right)^{-1}-\hat{B}\right)^{-1} \frac{\partial f^0}{\partial\epsilon}\ \mathrm{d}\epsilon}
	\end{equation}
	As in the case of the conductivity tensor, this result is equivalent to the result of Ref.~\cite{Mikhail1980} (equation~8b) except for the sign of $\hat{B}$.

	\subsection{Landau quantization}
	\subsubsection{Conductivity tensors}
	In order to include Landau quantization, we start from equation~\eqref{app:eq:start_Landau} and replace $\hbar^2 k_i^2 m_{be_{ii}}^{-1}$ by $2(\gamma^\ast-j\hbar\omega_c)$ (in the supplement, we use $\gamma^\ast$ as a short form for $\gamma (\epsilon^\ast)$). We get
	\begin{equation}
		\mathbf{j} = -e\int \frac{2(\gamma^\ast-j\hbar\omega_c)}{{\gamma'}^\ast} 
		\left(\left(\frac{e}{{\gamma'}^\ast}\hat{m}_{be}^{-1}\hat{\tau}\right)^{-1}-\hat{B}\right)^{-1} \frac{\partial f^0}{\partial\epsilon}\mathbf{E} D(\epsilon)\ \mathrm{d}\epsilon.
	\end{equation}
	The central step is now to put in the density of states describing Landau quantization (cf.\ Eq.~\eqref{eq:DOS_Landau}) instead of the purely semiclassical density of states:
	\begin{equation}
		\mathbf{j} = {}  \frac{(2m_{be,\parallel})^{1/2} \abs{eB} e}{2\pi^2\hbar^2} \sum_{s=\pm 1/2} \sum_{n=0}^{\infty} \int (\gamma^\ast-j\hbar\omega_c)^{1/2}
		\left(\left(\frac{e}{{\gamma'}^\ast}\hat{m}_{be}^{-1}\hat{\tau}\right)^{-1}-\hat{B}\right)^{-1} \frac{\partial f^0}{\partial\epsilon}\ \mathrm{d}\epsilon\ \mathbf{E} 
	\end{equation}
	As discussed below, the changed density of states enters also the relaxation time $\hat{\tau}$. With $\mathbf{j} = \hat{\sigma} \mathbf{E}$, we obtain for the electrical conductivity tensor
	\begin{equation}
		\hat{\sigma} = \sum_{\mathrm{pockets}}\frac{(2m_{be,\parallel})^{1/2} \abs{eB} e}{2\pi^2\hbar^2} \sum_{s=\pm 1/2} \sum_{n=0}^{\infty} \int (\gamma^\ast-j\hbar\omega_c)^{1/2}
		\left(\left(\frac{e}{{\gamma'}^\ast}\hat{m}_{be}^{-1}\hat{\tau}\right)^{-1}-\hat{B}\right)^{-1} \frac{\partial f^0}{\partial\epsilon}\ \mathrm{d}\epsilon. \label{app:eq:sigma_Landau} 
	\end{equation}
	The thermoelectric conductivity tensor can be derived by performing similar steps. It reads
	\begin{equation}
		\hat{\alpha} = \sum_{\mathrm{pockets}}\frac{(2m_{be,\parallel})^{1/2} \abs{eB}}{2\pi^2\hbar^2} \sum_{s=\pm 1/2} \sum_{n=0}^{\infty} \int \frac{\epsilon-\mu}{T} (\gamma^\ast-j\hbar\omega_c)^{1/2}
		\left(\left(\frac{e}{{\gamma'}^\ast}\hat{m}_{be}^{-1}\hat{\tau}\right)^{-1}-\hat{B}\right)^{-1} \frac{\partial f^0}{\partial\epsilon}\ \mathrm{d}\epsilon. \label{app:eq:alpha_Landau} 
	\end{equation}
	
	\subsubsection{Scattering time}
	
	As in the low-field case, the most critical point of the derivation is the scattering time. According to equation \eqref{eq:scattering_time_general}, it is inversely proportional to the density of states. In the purely semiclassical regime intravalley scattering is assumed, i.e.\ only the density of states of the respective pocket is inserted into the equation. Although the temperatures we target with this model are high enough to give rise to a non-negligible number of short wavelength phonons causing intervalley scattering \cite{Murray2007}, this assumption seems relatively well justified, because the three electron pockets are degenerated at zero field. Thus, it should not make a big difference for the energy dependence of the scattering time $\tau(\epsilon)$ if intervalley scattering is taken into account or not. In contrast, in the case of Landau quantization, the degeneracy of the valleys is lifted (cf.\ Fig.~\ref{fig:mu_nvsB}). Hence, the best way to describe $\tau(\epsilon)$ would probably be to also consider the electronic density of states of the other pockets convoluted with the density of states of the phonons contributing to intervalley scattering. However, this is out of the scope of this model, both conceptually and in terms of computation time. Therefore, two possible approximations were tried out and compared to experimental results:
	\begin{itemize}
		\item Only the density of states of the respective pocket was used (intravalley scattering). This implies a large difference between the mobilities of the pockets as the valleys' total electron density differs by a factor of up to 16 at $B=14\,T$.
		\item The total electronic density of states (normalized by a factor of three as to be in accordance with the low-field case) was inserted in equation \eqref{eq:scattering_time_general}. This is a rough approximation of intervalley scattering, because it neither takes into account the non-negligible energy of phonons taking part in intervalley scattering nor accounts for a possible difference in the scattering matrix element.
	\end{itemize}
	The second version was chosen, since it leads to a better agreement with the experimental results.

	\subsection{Observables}
	
	From the electrical conductivity tensor, the resistivity tensor $\hat{\rho}$ can be calculated by taking the inverse:
	\begin{equation}
		\hat{\rho} = \hat{\sigma}^{-1}
	\end{equation}
	The diagonal elements of $\hat{\rho}$ are the resistivities along the three axes and the off-diagonal elements give the Hall resistivities. The Seebeck effect of the system is calculated via
	\begin{equation}
		\hat{S}=\hat{\rho}\hat{\alpha},
	\end{equation}
	where the diagonal elements of $\hat{S}$ correspond to the Seebeck coefficients along the three crystallographic directions.
	
	\subsection{Computation}
	
	The predictions of the model described above were calculated with a python script. However, integrating diverging integrands is very time consuming, which is why several approximations had to be made in order to reduce the computing time. Firstly, for the hole valleys always the density of states without Landau quantization is used. This is justified, because the quantum limit of the holes at the $T$-point is in the vicinity of 40~T~\cite{Zhu2017} such that the way they are treated does not significantly influence the outcome. Secondly, the electron pockets are treated purely semiclassically when there are four or more Landau levels below $\mu-k_BT$. This criterion was adjusted such that the resulting curves are smooth at the transition between the semiclassical and the quantum treatment. Thirdly, the upper integration limit is set to the minimum of $\mu+20k_BT$ and $\mu+170\ \mathrm{meV}$. The first value, which could seem very high, is required at low temperatures in order to make the results for the Hall effect converge. The latter is used, because the Lax model is not valid any more above this energy \cite{Liu1995,Fuseya2015}. Despite these approximations, the calculations were very slow in some parts of the $(T,B,\Theta)$-space. In this case, the results were linearly interpolated leading to kinks in a few of the shown curves.
	
	\section{Supplementary data}
	
\begin{figure*}
		\centering
		\includegraphics[width=0.85\linewidth]{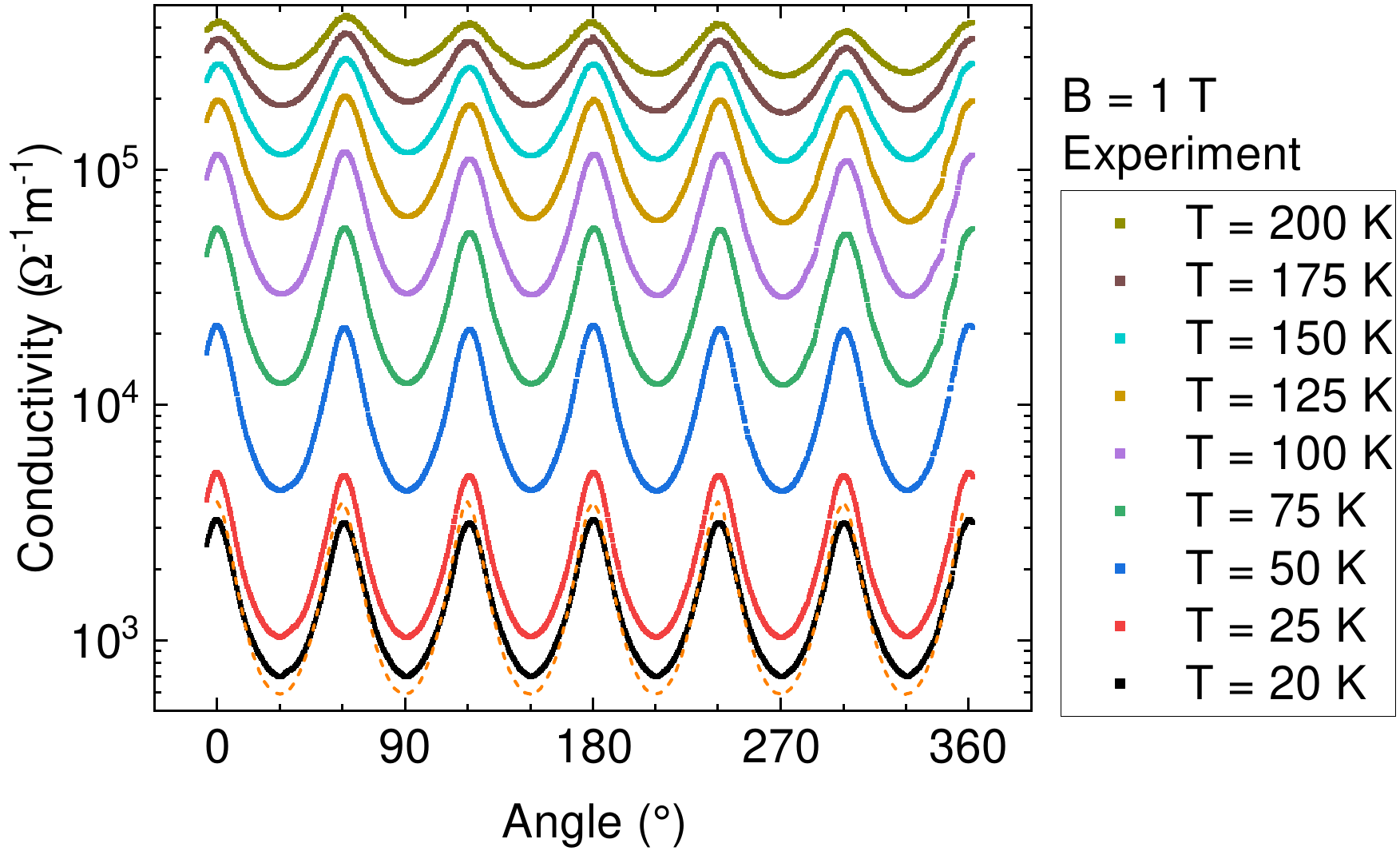}
		\caption{Electrical conductivity ($\sigma_{zz} \approx 1/\rho_{zz}$ because of compensation) as a function of the orientation of the magnetic field at 1~T and several temperatures between 20~K and 200~K. The dashed orange curve shows the theoretical result for $T=20$~K to allow for better comparison with the experimental data.}
		\label{fig:SigmavsTheta}
	\end{figure*}

\begin{figure*}
		\centering
		\includegraphics[width=0.85\linewidth]{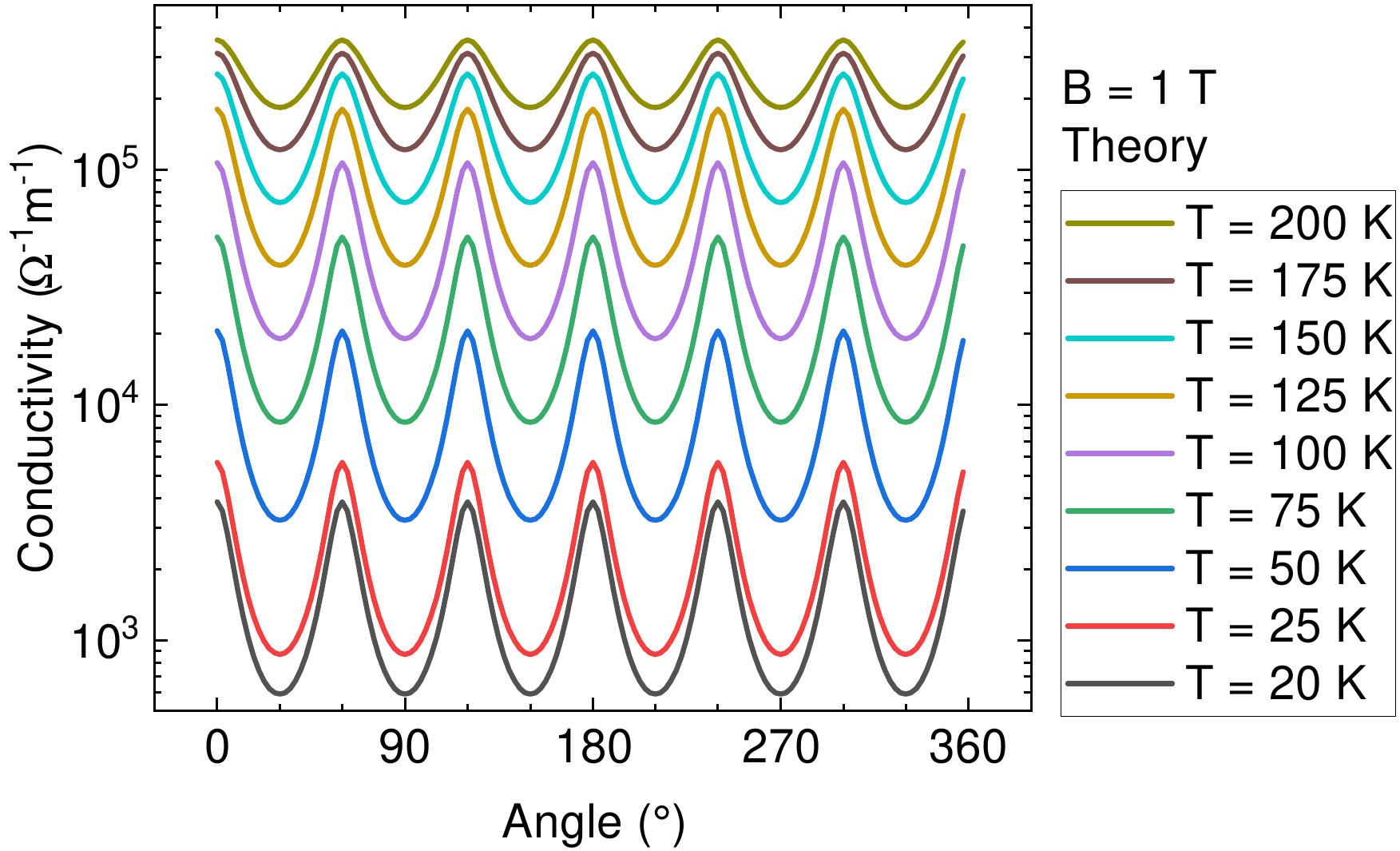}
		\caption{Theoretical prediction of the electrical conductivity $\sigma_{zz}$ as a function of the orientation of the magnetic field at 1~T and several temperatures between 20~K and
			200~K. Theoretical and experimental results are in very good agreement with each other. Landau quantization was not taken into account when simulating these curves.}
		\label{fig:SigmavsTheta_theory}
	\end{figure*}

\begin{figure*}
		\centering
		\includegraphics[width=1\linewidth]{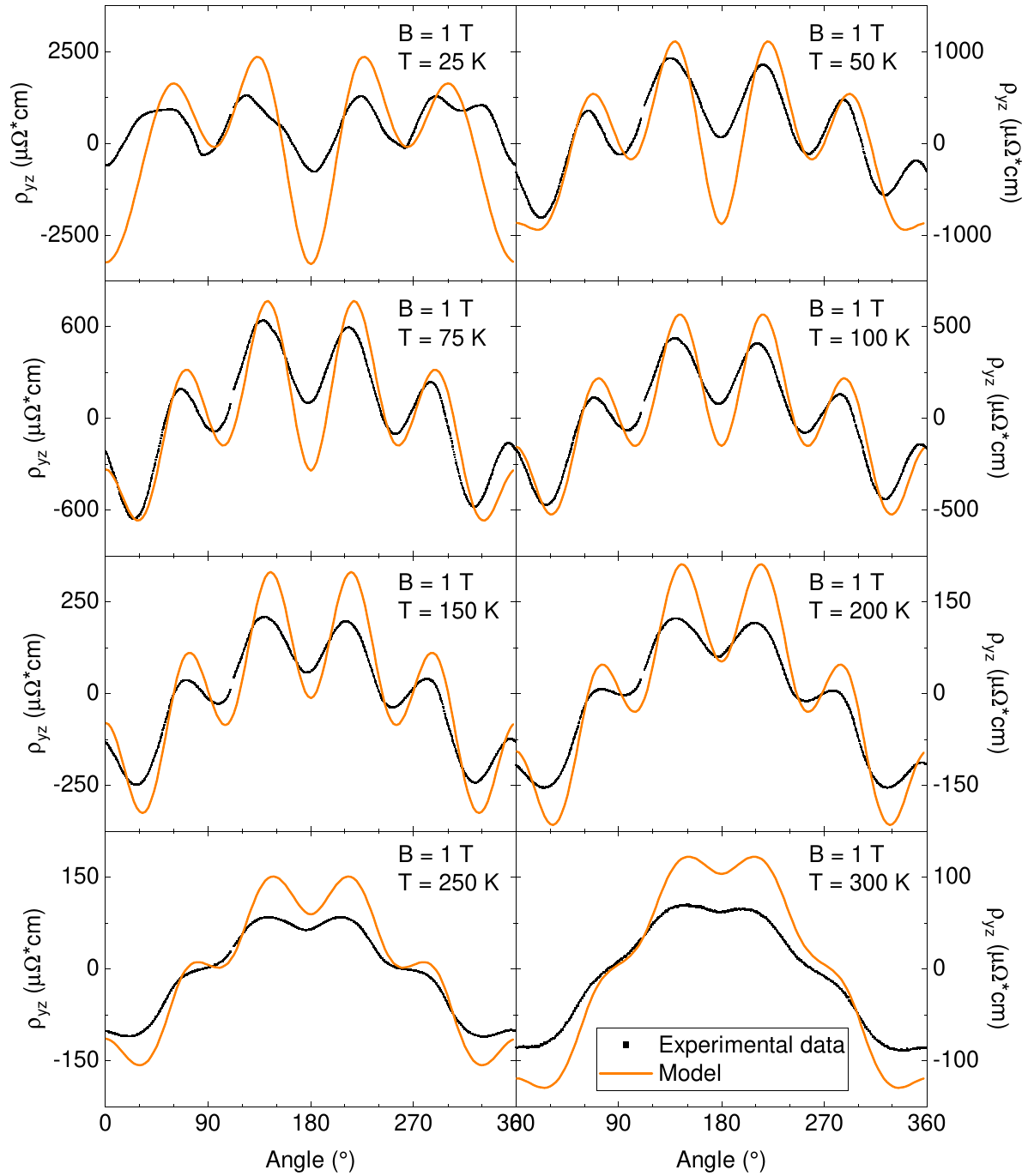}
		\caption{Hall resistivity $\rho_{yz}$ as a function of the orientation of the magnetic field at 1~T and several temperatures between 25~K and 300~K. Experimental data is shown in black and compared with theoretical data (orange). Landau quantization was not taken into account when simulating these curves.}
		\label{fig:Hall}
	\end{figure*}

	Figure~\ref{fig:SigmavsTheta} shows angle-dependent magnetoconductivity $\sigma_{zz}$ at 1~T and several temperatures between 20~K and 200~K. Figure~\ref{fig:SigmavsTheta_theory} shows the corresponding theoretical curves. Figure~\ref{fig:Hall} shows the angular dependence of the Hall resistivity $\rho_{yz}$ at 1~T and several temperatures between 25~K and 300~K as well as the corresponding theoretical curves.

	\clearpage
	
\end{document}